\documentclass{aa}

\usepackage{graphicx}
\usepackage{txfonts}
\usepackage{natbib}

\bibpunct{(}{)}{;}{a}{}{,} % to follow the A&A style

\newcommand{\Mpc}{$h^{-1}$\thinspace Mpc}

\def\apj{ApJ}
\def\apjl{ApJL}
\def\apjs{ApJS}
\def\aj{AJ}
\def\aap{A\&A}
\def\mnras{MNRAS}

\begin{document}   
\sloppy

\title{Wavelet analysis of the cosmic web formation } 

\author{ J. Einasto\inst{1,2,3} \and G. H\"utsi\inst{1} \and E. Saar\inst{1,2} \and I. Suhhonenko\inst{1} 
 \and L. J. Liivam\"agi\inst{1}  \and M. Einasto\inst{1}  \and    V. M\"uller\inst{4}  \and A. A. Starobinsky\inst{5,6} \and E. Tago\inst{1}   \and E. Tempel\inst{1}}

\institute{Tartu Observatory, EE-61602 T\~oravere, Estonia
\and 
Estonian Academy of Sciences,  EE-10130 Tallinn, Estonia
\and 
ICRANet, Piazza della Repubblica 10, 65122 Pescara, Italy
\and
Leibniz-Institut f\"ur Astrophysik Potsdam, An der Sternwarte 16, D-14482 Potsdam,
  Germany
\and
Landau Institute for Theoretical Physics, Moscow 119334, Russia
\and
Research Center for the Early Universe (RESCEU), Graduate School of Science,
The University of Tokyo, Tokyo 113-0033, Japan
}

\date{ Received 4 November2010/ Accepted 5 May 2011} 

\authorrunning{J. Einasto et al.}

\titlerunning{Formation of the cosmic web}

\offprints{J. Einasto, e-mail: einasto@aai.ee}
\abstract 
% context
{According to the modern cosmological paradigm, galaxies and galaxy systems form
  from tiny density perturbations generated during the very early phase of the
  evolution of the Universe.  } 
% aims
{Using numerical simulations, we study the evolution of the density perturbation phases of different scales to understand the formation and evolution of the cosmic web.}
% methods
{We apply the wavelet analysis to follow the evolution of high-density regions (clusters and superclusters) of the cosmic web.} 
% results
{We show that the maxima and minima positions of  density waves (their spatial phases) almost do not change during the evolution of the structure.  Positions of density perturbation extrema of are  more stable for  large scale perturbations. In the context of the present study we call density waves of scale $\geq 64$~\Mpc\  large, waves of scale $\simeq 32$~\Mpc\  medium, and waves of scale $\simeq 8$~\Mpc\  small, within a factor of 2. }
% conclusions
{In the cosmic structure formation of the synchronisation (coupling) of density waves of different scales plays an important role.}

\keywords{cosmology: large-scale structure of the Universe;   cosmology: early Universe; cosmology: theory; methods:numerical}

\maketitle

\section{Introduction}

The basic structural elements of the Universe are filamentary superclusters and voids forming a web-like structure -- the supercluster-void network \citep{{Einasto:1980}, {Zeldovich:1982}, {de-Lapparent:1986}, {Bond:1996}}.
 The standard cosmological paradigm predicts that a period of accelerated expansion, dubbed inflation \citep{Starobinsky:1980rt,Guth:1981ys,Linde:1982fr,Albrecht:1982}, generated density fluctuations \citep{Mukhanov:1981,Hawking:1982,Starobinsky:1982zr,Guth:1982} as well as primordial gravitational waves \citep{Starobinskii:1979vn} through quantum-gravitational processes.  In the simplest form of this scenario, the primordial density field is predicted to form a statistically homogeneous, isotropic and almost-Gaussian random field, after the transition from quantum to approximate classical description of perturbations. 

If the hypothesis of primordial Gaussianity is correct, then density waves of different scales began with random and uncorrelated spatial phases. As the density waves evolve, they interact with others in a non-linear way. This interaction leads to the generation of non-random and correlated phases which form a spatial pattern of the present cosmic web.  Owing to non-linear processes during galaxy formation and the physical biasing problem (almost no galaxies form in low-density regions), the present density field is highly non-Gaussian. There have been a number of attempts to gain quantitative information on the behaviour of phases in gravitational systems; for a review see \citet{Coles:2009} and references therein. 

Using numerical simulations, \citet{Ryden:1991fk} showed that initial phase information is rapidly lost in short wavelengths during evolution. \citet{Hikage:2005bd} analysed the clustering of SDSS galaxies using the distribution function of the sum of Fourier phases. Fourier phases are statistically independent of the Fourier amplitudes, thus the phase statistics plays a complementary role to the conventional two-point statistics of galaxy clustering. Gaussian fields have a uniform distribution of the Fourier phases over $0\leq \theta_k \leq 2\pi$. Therefore, characterising the correlation of phases is expected to be a direct means to explore non-Gaussian features.  From the comparison of observations with mock catalogues constructed from N-body simulations, the authors find that the observed phase correlations for the galaxies agree well with those predicted by the spatially flat $\Lambda$CDM model, evolved from Gaussian initial conditions.

The analysis in Fourier space is, however, not sensitive to the location of particular high-density features in real space, such as filaments, clusters, and superclusters. To have a better understanding of the texture of the cosmic web, the web must be studied in the real space. Different statistical measures have been used to describe quantitatively the cosmic texture, for recent  reviews see \citet{Martinez:2002ye},  \citet{Saar:2009}, and \citet{van-de-Weygaert:2009bh}.

One of these statistics is the wavelet analysis, which analyses properties of waves of various scales in real space \citep[see][and references therein]{Jones:2009dq}.  Wavelet analysis has been used to detect voids and filaments in the Center for Astrophysics (CfA) survey first slice \citep{Slezak:1993nx}, to de-noise the galaxy distribution \citep{Martinez:2005kl}, to detect  the integrated Sachs-Wolfe effect in the cosmic microwave background (CMB) radiation \citep{Vielva:2006tg}, {to study the discreteness effects in simulations \citep{Romeo:2008}}, 
and for many other purposes where the spatial position of structural elements is important.

The goal of this paper is to investigate the evolution of the texture of the cosmic web.  It is well known that on small scales the original phase information is lost during the non-linear stage of the evolution \citep{Ryden:1991fk}.  On the other hand, the main large-scale skeleton of the texture of the cosmic web is already determined by the initial gravitational potential field \citep{Kofman:1988}.  {If considered from the point of view of the Fourier decomposition, maxima and minima in {\em any} spatial distribution occur in the points where phases of the Fourier modes are synchronised. However, there is a non-trivial problem, which is not completely solved yet.

  In the classical ``down-top'' model, like the isocurvature model, there are no built-in large-scale features. The structure formation starts from small-scale systems, which grow by random clustering.  In the classical ``top-down'' model, like the adiabatic neutrino-dominated hot dark matter model, there is a built-in cut-off scale, which determines the scale of the structure.  The presently accepted dark energy dominated $\Lambda$CDM model is essentially a ``down-top'' model, because the structure formation starts from small systems.  This model differs from the classical ``down-top'' model in one important detail -- here a broad power spectrum of density perturbations is present.  Objects of a smaller scale and mass form earlier.  But in the case of a broad power spectrum of perturbations, it becomes a non-trivial question whether extrema of perturbations of a given scale will remain at the same places if perturbations of larger scales are added. This may occur only if some phase synchronisation or coupling between perturbations of {\em different scales} exists.  In the case of the broad wave spectrum, extrema of density perturbations should define locations where gravitationally bound objects and voids form first.  On the other hand, the gravitational potential defines the location of the skeleton of the cosmic web knots.  Consequently, it is not clear at all why extrema of density perturbations coincide with knots defined by the gravitational potential.  In other words: Why is the skeleton  stable in the ``down-top'' $\Lambda$CDM model?  As we see in this paper, just because of this synchronisation between waves of different scales.}

Studies of Fourier phases show that the phase coupling in the non-linear regime plays an important role in the formation of the fine texture of the cosmic web \citep{Chiang:2002}.   To avoid complications caused by the highly non-linear regime, we concentrate on the evolution of waves at intermediate and large scales using the wavelet decomposition of the evolving density field.
The Fourier modes are fully specified by their wavelength, their orientation, and phase. Because the phase determines where the maxima and minima are located along a Fourier mode, we also use the same terminology (somewhat less strict this time) once we speak about the wavelet decomposition of the density field. Here we assume that the (spatial) phase and the locations of the maxima and minima carry the same information, and thus will use these terms interchangeably in the following. Also, quite often the locations of the cells inside the cubical density grid are located as $(i,j,k)$, with $i$, $j$, and $k$ are integers that run from $1$ to $N$, where $N=256$ is most often assumed throughout the work.

We shall focus our attention on the two main problems: the evolution of phases (positions of maxima) of density perturbations at medium and large scales, and the phase coupling (synchronisation) of perturbations of different scales.  
To follow the evolution of perturbations of different size, we use simulations in boxes of various sizes from 100 to 768 ~\Mpc. To find the sensitivity of our results to the resolution, we make simulations with $256^3$ and $512^3$ cells and equal number of particles. For comparison with the real Universe, we shall calculate the density field and its wavelet decompositions for a slice (wedge) of the Sloan Digital Sky Survey (SDSS).  Preliminary results of this study have been reported at several conferences \citep{Einasto:2006mz,Einasto:2006zr,Einasto:2009ly}.  This paper is a follow-up of a study by \citet{Einasto:2005a} of the environmental effects of clusters in SDSS and simulations. 

In the next section we describe the  numerical models used in this study. We also make a qualitative  wavelet analysis of  the simulated density field,  follow the density evolution in time, and compare the evolution of density waves of various scales. In Section 3 we make a correlation analysis of wavelet-decomposed density fields. In Section 4 we analyse the luminosity density field of the SDSS, and study the role of phase synchronisation in the formation of clusters and superclusters. In Section 5 we discuss our results. The last section  gives our main conclusions.

\section{Qualitative wavelet analysis of the cosmic web evolution }

\subsection{Modelling the  cosmic web evolution }

In order to understand the evolution of the supercluster-void phenomenon, numerical simulations need to be performed in a box that contains both small- and large-scale waves.  The most common systems of galaxies are groups and galaxy clusters with a characteristic scale of $\sim 1$~\Mpc, therefore the simulation must have at least a resolution of this scale.  On the other hand, the largest non-percolating systems are superclusters, which have a characteristic scale up to $\sim 100$~\Mpc.  Superclusters have a very different richness from small systems like the Local supercluster to very rich systems like the Shapley supercluster  \citep{Einasto:2001ff}.  Clearly this variety has its origin in density perturbations of still larger scales.  Thus,  to understand the supercluster-void phenomenon correctly, the influence of very large-scale density perturbations should be studied, too.

\begin{figure*}[ht]
\centering
\resizebox{0.9\textwidth}{!}{\includegraphics*{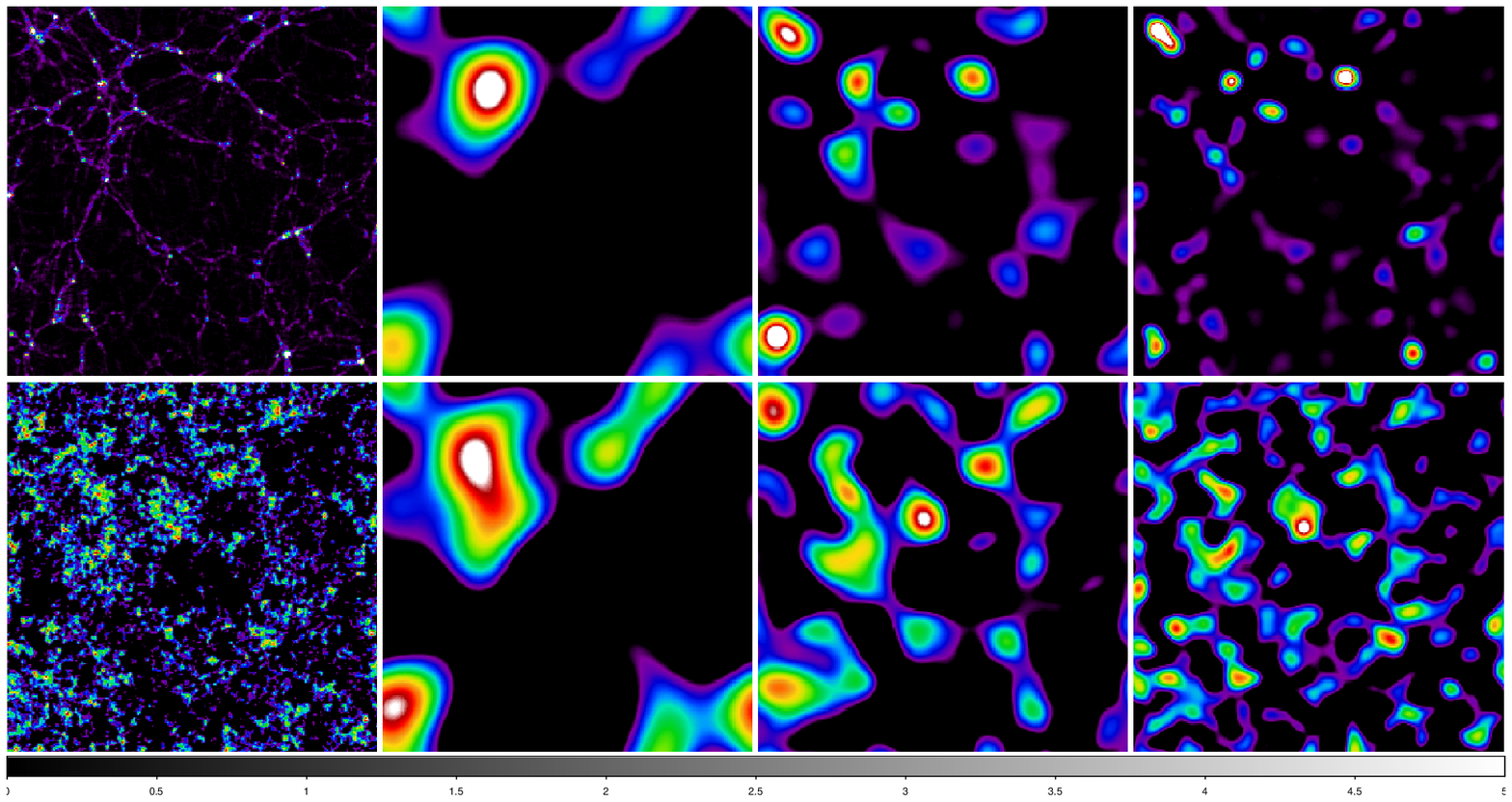}}
\resizebox{0.9\textwidth}{!}{\includegraphics*{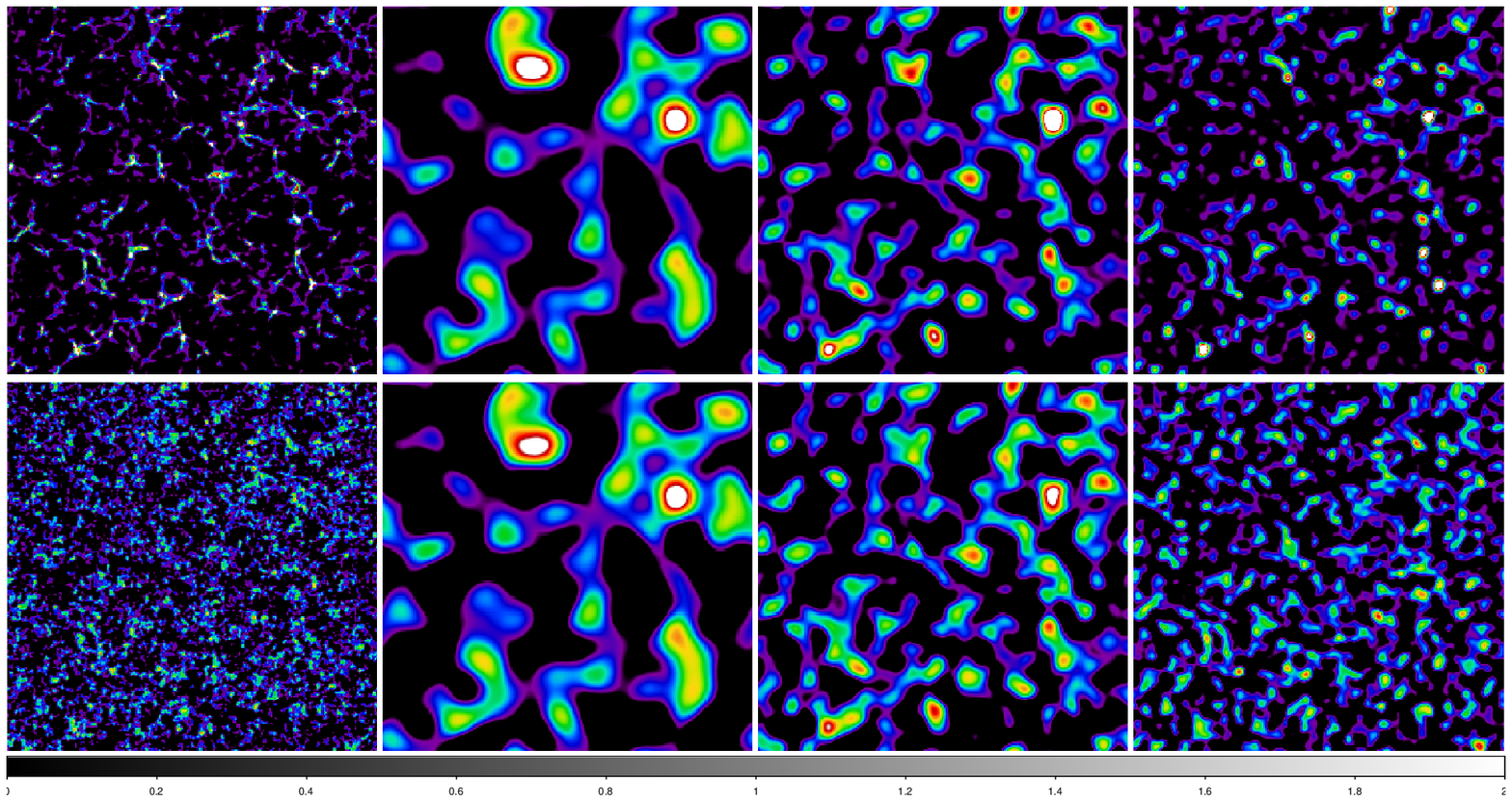}}
\caption{High-resolution  density fields and their wavelet decompositions for models M100 and M768, upper two and lower two rows, respectively. Model M100 fields are given  at $k=1$ coordinate, M768 fields  at $k=217$ coordinate.  Left panels show the density fields, next panels their wavelet decompositions. For both models upper panels  show the field and wavelets at redshift $z = 0$, and lower panels at  redshift $z = 100$. For model M100 we show the wavelets $w6$, $w5$, and $w4$, for model M768 wavelets $w5$, $w4$, and $w3$. 
In plotting density fields and wavelets we use only over-dense regions: i.e., lower density limit (which corresponds to black level in DS9 colour palette SLS) is taken equal to 1 for the density field, and 0 for wavelets. Upper density levels  (which correspond to white in palette SLS) are given in Table~\ref{tab:wavdata}.   Table~\ref{tab:wavdata} gives also the upper smoothing wavelength Sw$n$d in \Mpc\  for all wavelets.  }
\label{fig:m100wav}
\end{figure*}

{\scriptsize
\begin{table}[ht]
%\centering
\caption{Parameters used in wavelet figures.}
\begin{tabular}{rrrrrrr} 
\hline\hline 
\noalign{\smallskip}
Model & $z$       & DFu & Dw6u & Dw5u & Dw4u  & Sw$n$u  \\
(1)       & (2)          & (3)   &  (4)  & (5)    & (6)    & (7)      \\
\noalign{\smallskip}
\hline
 \noalign{\smallskip}
M100  & 0            & 40     & 1.0     & 1.8       & 5.0  &  50 \\
M100  & 100        & 1.15   & 0.0085  & 0.018  & 0.03&   50\\
\\
M256  & 0            & 20   & 0.55    & 1.5     & 3.5  &    128\\
M256  & 1            & 10   & 0.3     & 0.75  & 1.75 &     128\\
M256  & 5            & 4      & 0.09    & 0.2   & 0.45 &     128\\
M256  & 10          & 2      & 0.05   & 0.1    & 0.2 &     128\\
\\
M768  & 0            & 10    & 0.25    & 0.86   & 2.0  &    192\\
M768  & 100        & 1.08  & 0.004 & 0.009  & 0.018 &     192\\
\noalign{\smallskip}
\hline
\label{tab:wavdata}                                                                                    
\end{tabular}
\tablefoot{
\\
\noindent 1: Model;\\
\noindent 2: Redshift $z$;\\
\noindent 3: Upper density limit of the high-resolution density field used in 
Figs.\ref{fig:m100wav}, \ref{fig:m256wav};\\
\noindent 4: Upper density limit of the wavelet $w6$ ($w5$ for the model M768) used in Figs.\ref{fig:m100wav}, \ref{fig:m256wav};\\
\noindent 5: Upper density limit of the wavelet $w5$ ($w4$) used in Figs.\ref{fig:m100wav}, \ref{fig:m256wav};\\
\noindent 6: Upper density limit of the wavelet $w4$ ($w3$) used in Figs.\ref{fig:m100wav}, \ref{fig:m256wav}; \\
\noindent 7: Upper smoothing wavelength in \Mpc\ of the largest wavelet Sw6u (Sw5u);
 the upper smoothing wavelength of the next  wavelet Sw5u (Sw4u) is 2 times smaller than that of Sw6u  (Sw5u), and that of Sw4u (Sw3u) is 4 times smaller than that of Sw6u  (Sw5u). Lower smoothing wavelengths of all the wavelets are 2 times smaller than  the upper ones. }
\end{table}
}

 To have both high spatial resolution and the presence of density perturbations in a broad interval of scales, we  used in this analysis simulations in boxes of sizes 100~\Mpc, 256~\Mpc, and 768~\Mpc, with the resolution of $256^3$  particles and simulation cells; these models are designated as M100, M256 and M768, respectively. For simulations we used the AMIGA code  \citep{Knebe:2001qa}.  This code uses an adaptive mesh technique in regions where the density exceeds a fixed threshold.  For comparison we used a model of box size 256~\Mpc\ with the resolution of $512^3$ particles and cells, this model is designated L256. In the last case we used the GADGET code for simulations \citep{Springel:2005a}.  Results obtained from models with different resolution are very similar in the context of the present study, thus in the graphical representation of our results we mostly use models M100, M256, and M768. The model L256 has been used to find a catalogue of density field (DF) clusters of galaxies,  to study the dependence of the mass of DF clusters on the density of the environment, and to compare wavelets of models with different cut-off scale (see Figure~\ref{fig:L256w1} below).

The initial density fluctuation spectrum was generated using the COSMICS code by \citet{Bertschinger:1995}\footnote{\tt http://arcturus.mit.edu/cosmics}.  We assumed cosmological parameters $\Omega_{\mathrm{m}} = 0.28$, $\Omega_{\Lambda} = 0.72$, 
$\sigma_8 = 0.84$, and the dimensionless Hubble constant $h = 0.71$; to generate the initial data we took the baryonic matter density $\Omega_\mathrm{b}= 0.044$.  Calculations started in an early epoch, $z=500$.  Particle positions and velocities were extracted for 13 epochs between redshifts $z=100, \dots, 0$. In the present study we used only a part of these data, depending on the goal of the task.

\begin{figure*}[ht]
\centering
\resizebox{0.9\textwidth}{!}{\includegraphics*{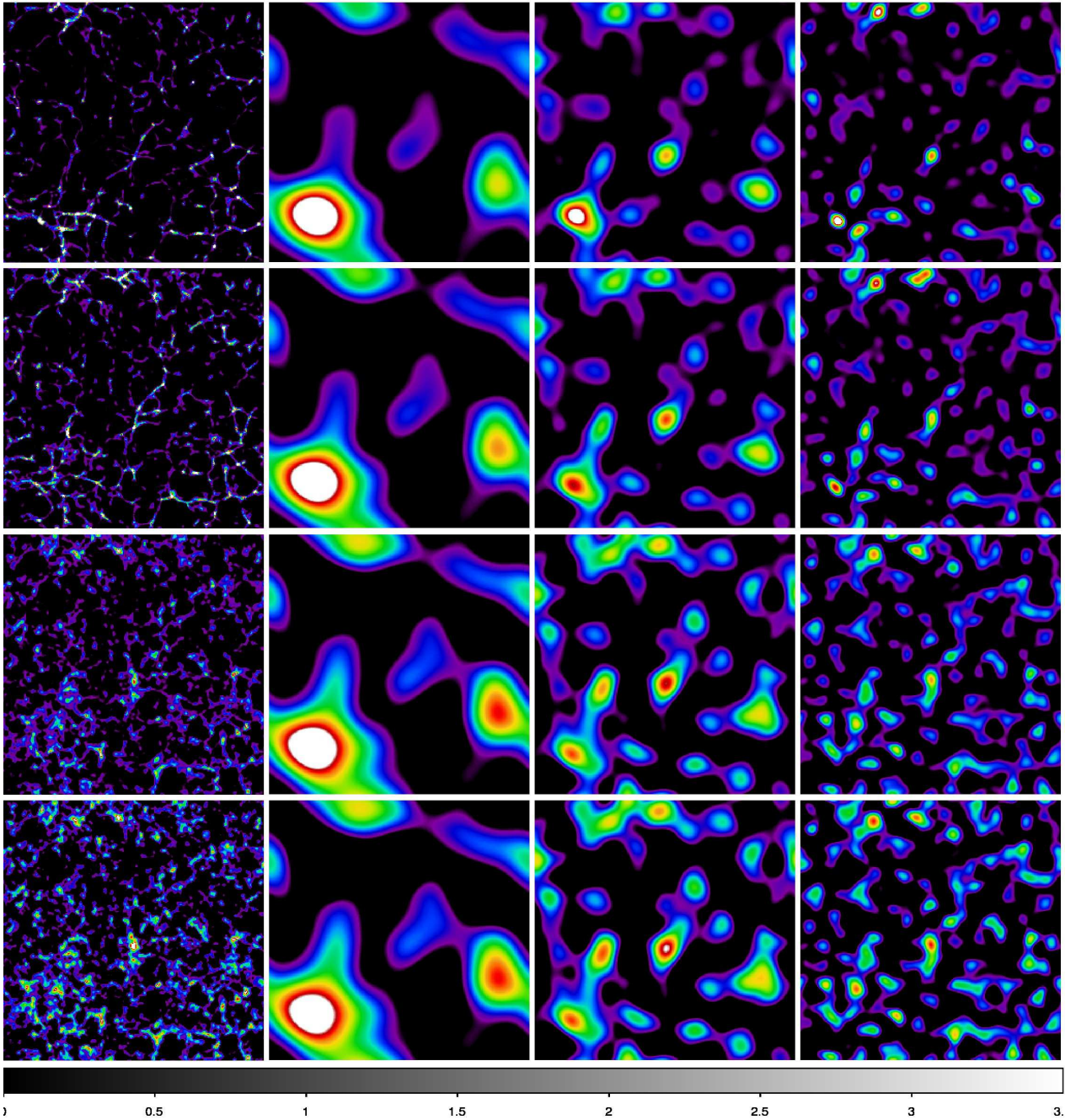}}
\caption{High-resolution density field of the model M256 is shown in the left column, at $k=153$ coordinate. The second, third, and fourth columns shows the  wavelet $w6$, $w5$, and $w4$ decompositions at the same $k$, respectively. The upper row gives data for present epoch, $z=0$, the second row for redshift $z=1$, the third row for redshift $z=5$, and the last row for redshift $z=10$. Densities are expressed in linear scale.  Parameters used for plotting are given in  Table~\ref{tab:wavdata}. 
 }
\label{fig:m256wav}
\end{figure*}

\subsection{Wavelet decomposition of simulated density fields}

To see the effects of density waves of different scales and to understand the evolution of the density field, we shall use our models M100,  M256, and M768.  The density fields were calculated for all simulation epochs and were used to find the wavelets up to the level 6.  The analysis was made in three dimensions. For illustrations of the results we use two-dimensional slices in $i,~j-$coordinates through the simulation box at fixed $k-$coordinates. The calculation of wavelets is explained in  Appendix A.3.

The evolution of waves in  models M100 and M768 are shown in Fig.~\ref{fig:m100wav}.  High-resolution density fields at redshifts $z=0$ and $z=100$ are shown in the left panels  of Fig.~\ref{fig:m100wav}.  The wavelet decompositions of  these density fields are shown for the same slices in the following panels of the figure. We show three wavelet levels: $w6$, $w5$, and $w4$ for the model M100; and $w5$, $w4$, and $w3$ for the model M768.  These wavelets show the evolution of large and intermediate scale waves. Upper smoothing wavelengths and upper density levels for plotting are given in Table~\ref{tab:wavdata}. We note that the upper smoothing scale of a wavelet w$n$  is equal to the kernel diameter, used in the calculation of the density field $D_n$, see Appendix A.3.

To see the evolution of the density field  at intermediate time-steps we show in Fig.~\ref{fig:m256wav} the high-resolution  density fields of the model M256 at four redshifts:  $z = 10,~5,~1,~0$.  Wavelet decompositions at levels $w6$, $w5$, and $w4$ for the same redshifts are shown in the same Figure. Colour-coding of wavelets at different redshifts is chosen so that a certain colour  approximately corresponds to the density level, corrected by the linear growth factor for that redshift. Parameters used for plotting are given in the Table~\ref{tab:wavdata}.

\subsection{The evolution of density waves in time}

Let us first study the evolution of wavelets of various scales  in time. It is well known that during the early stage of the evolution of the Universe the main evolution is in the growth of amplitudes of density perturbations. In the early stage of the evolution, the growth of amplitudes is almost linear.

How well this approximation works in our numerical simulations can be seen when we compare wavelets of different scales at various redshifts. The largest wavelets shown in Figs.~\ref{fig:m100wav} and \ref{fig:m256wav} are $w6$ for the models M100 and M256, and $w5$ for the model M768.  Upper smoothing scales of these largest wavelets are 50, 128, and 192~\Mpc\ (models M100, M256, and M768, see Table~\ref{tab:wavdata}).  The upper smoothing scales of next level wavelets for these models are 25, 64, 96~\Mpc.  The upper smoothing scales of the lowest level wavelets, shown in Figs.~\ref{fig:m100wav} and \ref{fig:m256wav} for models M100, M256, and M768, are 12.5, 32, and 48~\Mpc, respectively.  The upper smoothing scales of the wavelet $w1$  are 1.56, 4, and 12~\Mpc\ for these models. 

A look at  Figs. \ref{fig:m100wav} and \ref{fig:m256wav} shows that in the model M100 there are already considerable changes in positions and shapes of density configurations in the wavelet $w5$ (upper smoothing scale 25~\Mpc). In the model M256 changes of the wavelet $w5$ (upper smoothing scale 64~\Mpc) are still relatively small. 

In the  lowest scale wavelets ($w4$ of the model M256) the patterns of density distributions at various epochs are still fairly similar, however, changes in positions and density levels are more visible. These scales are upper limits of wavelengths of given wavelets, lower limits are twice smaller. Thus the mean smoothing scale of the  wavelet $w4$ of the model M256  is $32/\sqrt{2} = 22.6$~\Mpc.  Changes of patterns in the wavelet $w4$ of the model M100 (mean smoothing scale 8.8~\Mpc) are already fairly strong, both in the position and the density level.

\begin{figure}[ht]
\centering
\resizebox{0.49\textwidth}{!}{\includegraphics*{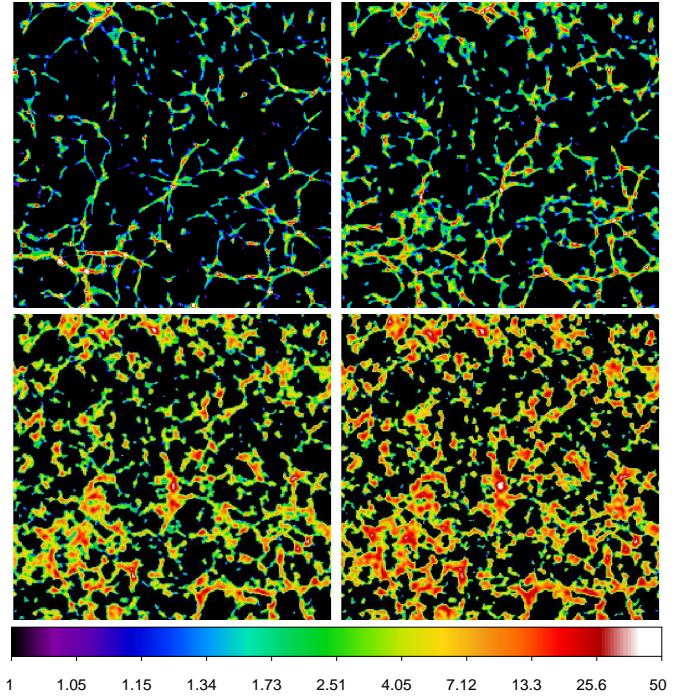}}
\caption{Evolution of the high-resolution density field of the model M256,  at $k=153$ coordinate. Upper panels are for  $z=0$ and $z=1$, lower panels for $z=5$ and $z=10$. The  densities are expressed in the logarithmic scale; this allows to see better the evolution in low-density regions.  Only over-densities are shown, i.e., lower density limit for plotting is taken 1. Upper densities for plotting with the DS9 package, corresponding to white, are 50, 25, 5, and 2.5 for redshifts 0, 1, 5, and 10, respectively. Colour codes shown at the bottom correspond to redshift $z=0$  (upper left panel). 
 }
\label{fig:m256.m0}
\end{figure}

To see the evolution of the filamentary density field  better, we show in Fig.~\ref{fig:m256.m0} the high-resolution density field of the model M256 again, with densities expressed in a logarithmic scale.  We use the same epochs and $k=153$ coordinate as in Fig. \ref{fig:m256wav}.  In the lower left region a very rich supercluster is located, visible in the wavelet $w6$ of Fig. \ref{fig:m256wav} as a large white area. The position of the supercluster does not change with time, the density contrast increases approximately proportional to the linear growth factor (used in setting colours for plotting). Most visible effects are the contraction of the filamentary system towards the centre of the supercluster, thinning of filaments, and merging of small filaments. 

\begin{figure}[ht]
\centering
\resizebox{0.48\textwidth}{!}{\includegraphics{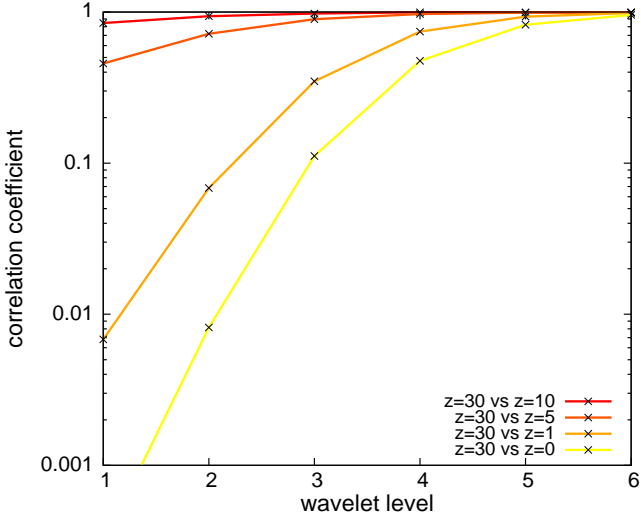}}
\caption{Behaviour of the correlation coefficient for different redshift pairs $(z_ji,z_j)\in \{(30,10);(30,5);(30,1);(30,0)\}$ for all  six wavelet levels. We see that for the largest smoothing scale, i.e. $w6$, all the correlators stay very close to $r=1$. At lower wavelet levels,  as the  redshift drops below $z=30$, the lines start to deviate from $r=1$. Thus, on the largest scales the information is approximately preserved, while on the smallest scales the information is gradually erased.}
\label{fig:correlate:fixed_w}
\end{figure}

Our simple qualitative analysis shows that the evolution of waves of medium and large scales  is approximately linear as expected.  Positions of maxima of these waves change very little with time. Waves of smaller scales change their positions of maxima, this can be considered as phase shifts of maxima. The shift is  larger for small-scale  waves. These shifts are caused by two effects:  the contraction of high-density regions (superclusters), and  the redistribution of matter on small scales, as illustrated in Fig.~\ref{fig:m256.m0}.  The contraction of the high-density regions is visible in Fig.~17 by 
\citet{Tempel:2009}, which shows the changes of the sizes of particle samples, giving rise to the present-day clusters of galaxies.  In central  regions of superclusters clusters formed by merging of a large number of primordial clusters, collected from an approximately spherical volume of  outer radius about 8~\Mpc. The lower the global density, the smaller is the radius of the sphere from which present clusters are formed.

\subsection{The comparison of evolution of density waves of various wavelengths}

Now we compare the evolution of density waves of different wavelengths, using the wavelet decomposition of our model density fields at various epochs. 

A look at Fig.~\ref{fig:m256wav} shows that at all redshifts high-density peaks of wavelets of {\em medium and large} scales almost coincide. In wavelets of  smaller scales, there is sometimes more than one peak in the high-density region of the next larger scale, but at least one peak is present there in practically  all cases. In other words, wavelets of various scales have a tendency of phase coupling or synchronisation in peak positions. Positions of high-density peaks of wavelets $w1$ and $w2$ coincide for any fixed redshifts (not shown in Figs.~\ref{fig:m100wav} and \ref{fig:m256wav}, but seen in original wavelet figures).  The highest of these small size density peaks also coincides in position with peaks of higher level wavelets.  We conclude that the growth of small-scale peaks is amplified in high-density regions of large waves: peaks in $w4$ to $w6$ in Fig.~\ref{fig:m256wav} and peaks of wavelets of various scale in Fig.~7 of \citet{Einasto:2011}. This amplification of density peaks of perturbations of various scales near peak positions is the reason for wave coupling (synchronisation).

The general conclusion  from the wavelet analysis of all scales and from the comparison with corresponding density fields is that the synchronisation of peak positions of wavelets of  {\em various scales} represents a general property of  the evolution of the density field of the Universe.  A quantitative analysis of this effect is given in the next section.

\begin{figure}[ht]
\centering
\resizebox{0.48\textwidth}{!}{\includegraphics{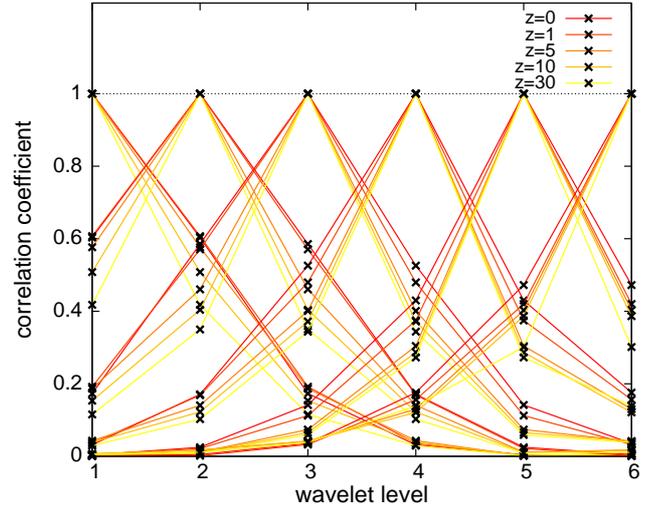}}
\caption{Correlators at fixed redshifts $z_i=z_j=0,1,5,10,30$ from top to bottom (from red to yellow) for high-density regions. Note that in this figure we define the high-density regions as the ones where the pixel values on the largest smoothed wavelet field reach  the top $10\%$ of all the values. Using these pixels, we define the mask, which is used for correlating only the highest density pixels with the corresponding values smoothed on a smaller scale.} 
\label{fig:correlate:fixed_z}
\end{figure}

\section{Correlation analysis of wavelet-decomposed density fields}

In this section we attempt to quantify some of the qualitative statements given above. To this end we perform the correlation analysis of wavelet-decomposed density fields. For clarity we will use only the model M256 because the other ones lead to very similar results. Here we have six wavelet levels: $w1, w2, \dots, w6$ with the upper smoothing scales of $4$, $8$, $\dots$, $128$ \Mpc, respectively. We use the simulated density fields at five different redshifts: $z=30$, $10$, $5$, $1$, $0$.

Most generally, we have chosen two redshifts, $z_i$ and $z_j$, along with two wavelet levels, $w_m$ and $w_n$, and calculated the following correlators:
\begin{equation}
r_{w_m z_i,\,w_n z_j}=\frac{\langle \delta_{w_m z_i}\delta_{w_n z_j}\rangle}{\sqrt{\langle\delta_{w_n z_j}^2\rangle\langle\delta_{w_n z_j}^2\rangle}}\,.
\label{eq:correlators}
\end{equation}
Here $\delta_{w_m z_i}$ corresponds to the wavelet-decomposed density field for level $w_m$ at redshift $z_i$. The angle brackets represent ensemble average, which under the ergodicity assumption is replaced by a simple spatial average. Note that all  wavelet-filtered fields have zero mean, i.e. $\langle\delta_{w_m z_i}\rangle=0$. Also note that we calculate the correlators for zero lag only, i.e., we do not shift one field with respect to the other one. In the following we will use two types of correlators:
\begin{enumerate}

\item Fixed wavelet scale correlators, i.e., $w_m=w_n$, at different redshifts.

\item Correlators at fixed redshifts, i.e., $z_i=z_j$, for different wavelet levels.

\end{enumerate}

In the first case we take one of the density fields always at redshift $z=30$, which is high enough for all  the scales of interest to be well in the linear regime. It is easy to understand that under the linear evolution, where the values of the density contrast $\delta$ get just multiplied by the same scale-independent but time-varying factor, the correlation coefficient should always stay at the value $r=1$, i.e., all  the initial information is well preserved. In Fig. \ref{fig:correlate:fixed_w} we show the behaviour of the correlation coefficient for different redshift pairs: $(z_ji,z_j)\in \{(30,10);(30,5);(30,1);(30,0)\}$ for all the six wavelet levels. It is easy to see that for the largest smoothing scale, i.e., $w6$, all the correlators stay very close to $r=1$, while later on, as the other redshift drops below than $z=30$, the lines start to decline from $r=1$, especially at the smallest scales. Thus, on the largest scales the information is approximately conserved, while on the smallest scales the information gets erased. The lower the redshift of the other density field, the higher the effect of the information erasure. In practice, for the cases $z=10$ and $z=5$ the loss of information is relatively modest for all wavelet levels. For $z=1$ and $z=0$ the information is approximately preserved if the wavelet level $w\ge 4$.

In Fig. \ref{fig:correlate:fixed_z} we plot the correlators at fixed redshifts $z_i=z_j=0,1,5,10,30$ from top to bottom (from red to yellow). For $w_m=1$ the curves are peaked at $w_n=1$, while they gradually  drop as $w_n$ is increased to higher values. Similarly, the curves for $w_m=2$ are peaked at $w_n=2$, and get reduced as the distance increases from this point. For the other $w_n$ values the behaviour is very similar. As long as the evolution proceeds in a linear manner, i.e., the growth  depends only on redshift, but is independent of the wavelet scale, the coupling kernels plotted in Fig. \ref{fig:correlate:fixed_z} should stay exactly the same. However, we see that the lower the output redshift, the broader are the coupling kernels, i.e., a nonlinear evolution in general leads to the additional coupling of the nearby wavelet modes. Also, note that in this figure we define the high-density regions as the ones where the pixel values on the largest scale wavelet field reach to the top $10\%$ of the values. Using these pixels we define the mask, which is used for correlating only the highest $10\%$ of the density values with the corresponding values smoothed  density on the smaller scale. Thus, in this case we have looked only how the highest density regions correlate with each other. We note that owing to nonlinear evolution, the coupling kernels become broader and broader with time. It is important to realise that even for the linear evolution of the Gaussian density field the nearby wavelet levels at fixed redshift are significantly coupled, because in this case the neighbouring levels tend to contain some of the common Fourier space modes. Assuming only the linear evolution, clearly  the coupling does not change with redshift.

\section{Wavelet analysis of the  SDSS luminosity density field}

\subsection{The  SDSS luminosity density field}

The luminosity density field was calculated using galaxy data of the Main sample of the contiguous Northern area of the SDSS data release 7 (DR7) \citep{Abazajian:2009kx}.
  The DR7 sample has, after applying extinction corrections, Petrosian $r$-magnitude limits, $14.5 \leq r \leq 17.77$.  We used for this analysis galaxies in the redshift interval $2700 \leq cz \leq 60000$ km~s$^{-1}$.  The SDSS data reduction procedure consists of two steps: (1) the calculation of the distance, the absolute magnitude, and the weight factor for each galaxy of the sample, and (2) the calculation of the luminosity density field using an appropriate kernel and a chosen smoothing length. We estimated total luminosities of groups and isolated galaxies in our flux-limited sample, using weights of galaxies that take into account galaxies and galaxy groups too faint to fall into the observational window of absolute magnitudes at the distance of the galaxy. When calculating luminosities of galaxies we regard every galaxy as a visible member of a group within the visible range of absolute magnitudes. For details of the calculation of the luminosity density field see  Appendix A.1. A supercluster catalogue based on the luminosity density field of SDSS DR7 is published by \citet{Liivamagi:2010}, and the luminosity function of galaxies by \citet{Tempel:2011}.

\begin{figure}[ht]
\begin{minipage}[h]{0.48\textwidth}
\centering
\resizebox{0.96\textwidth}{!}{\includegraphics*{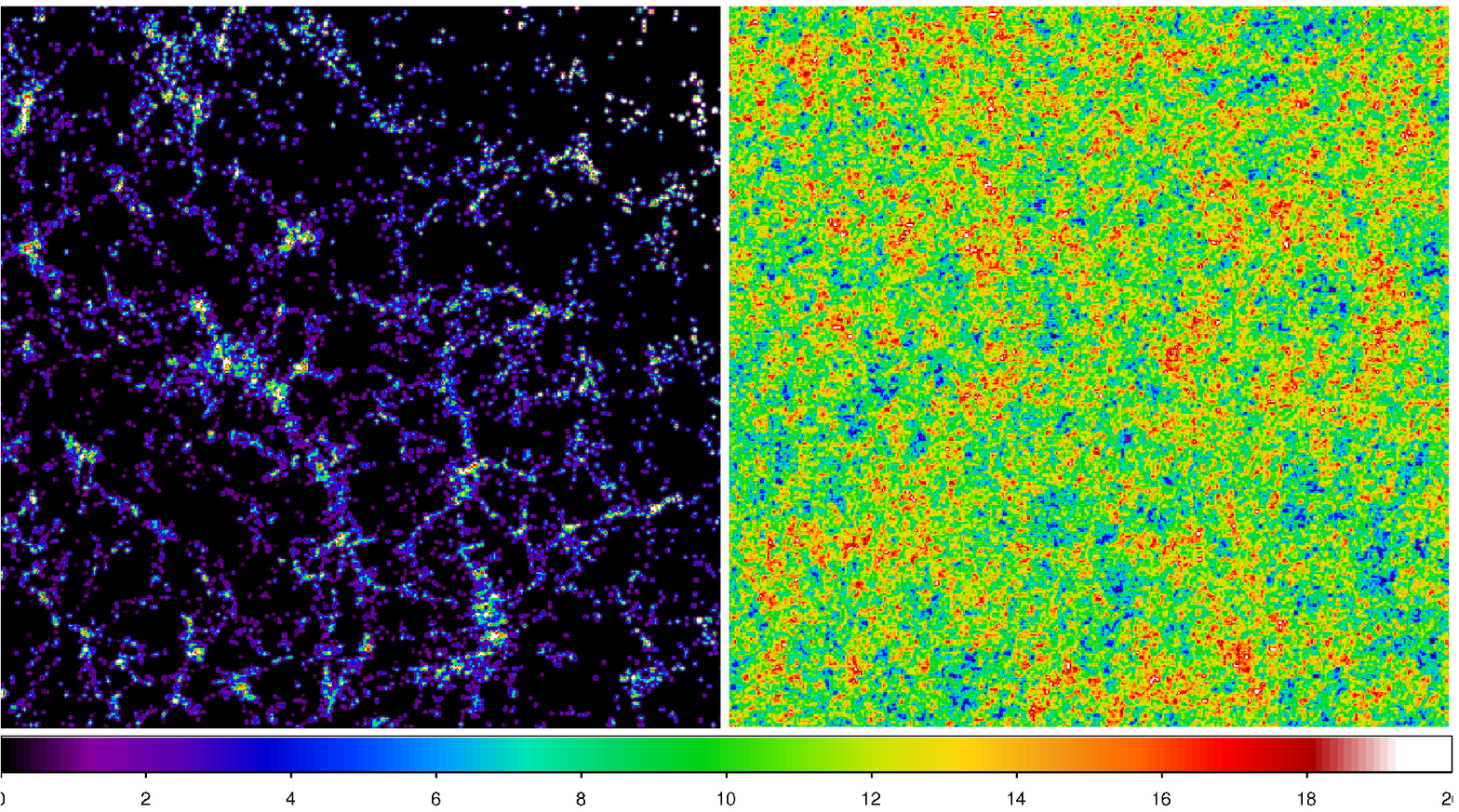}}\\
\resizebox{0.48\textwidth}{!}{\includegraphics*{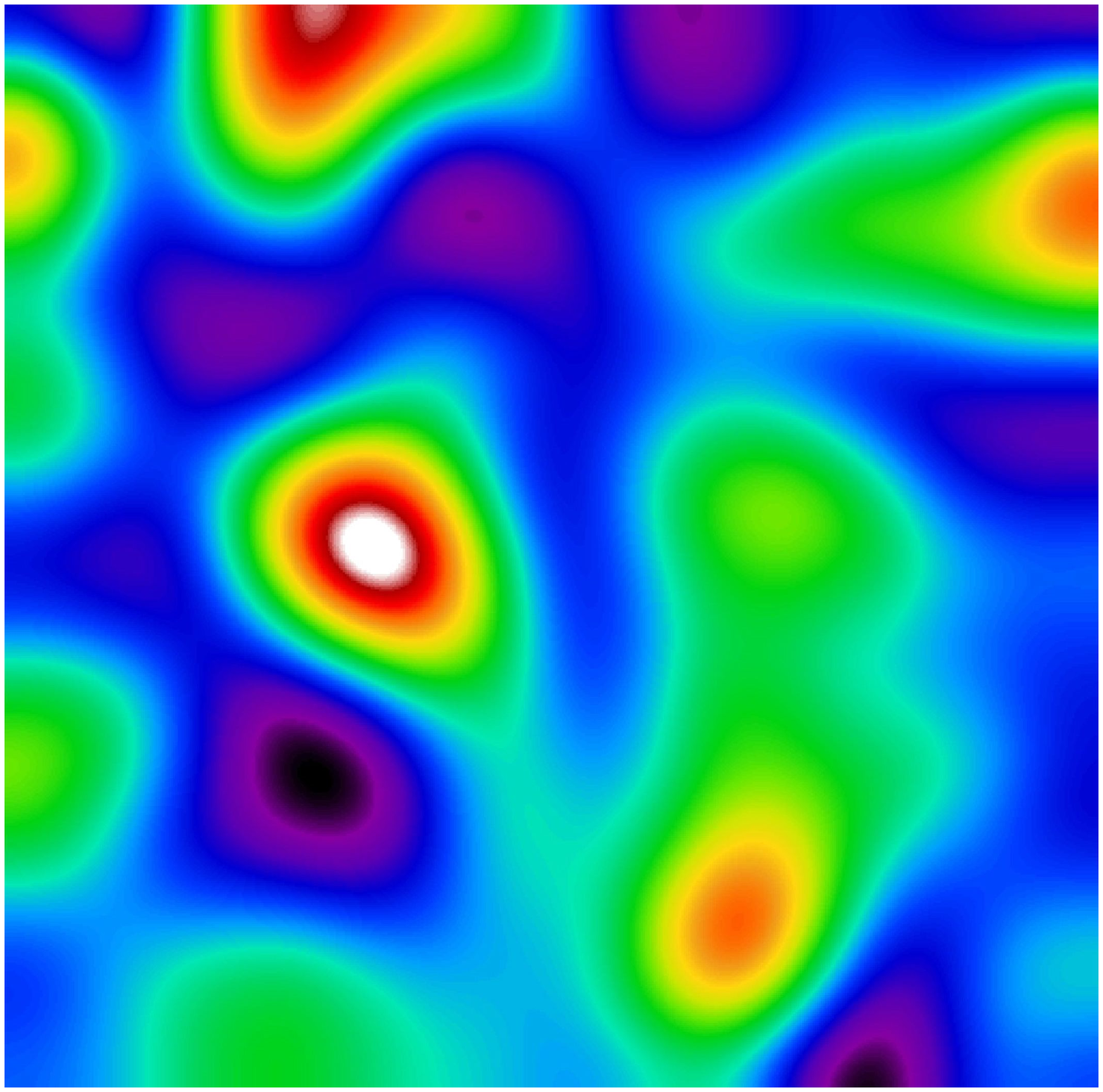}}
%\hspace{2mm} 
\resizebox{0.48\textwidth}{!}{\includegraphics*{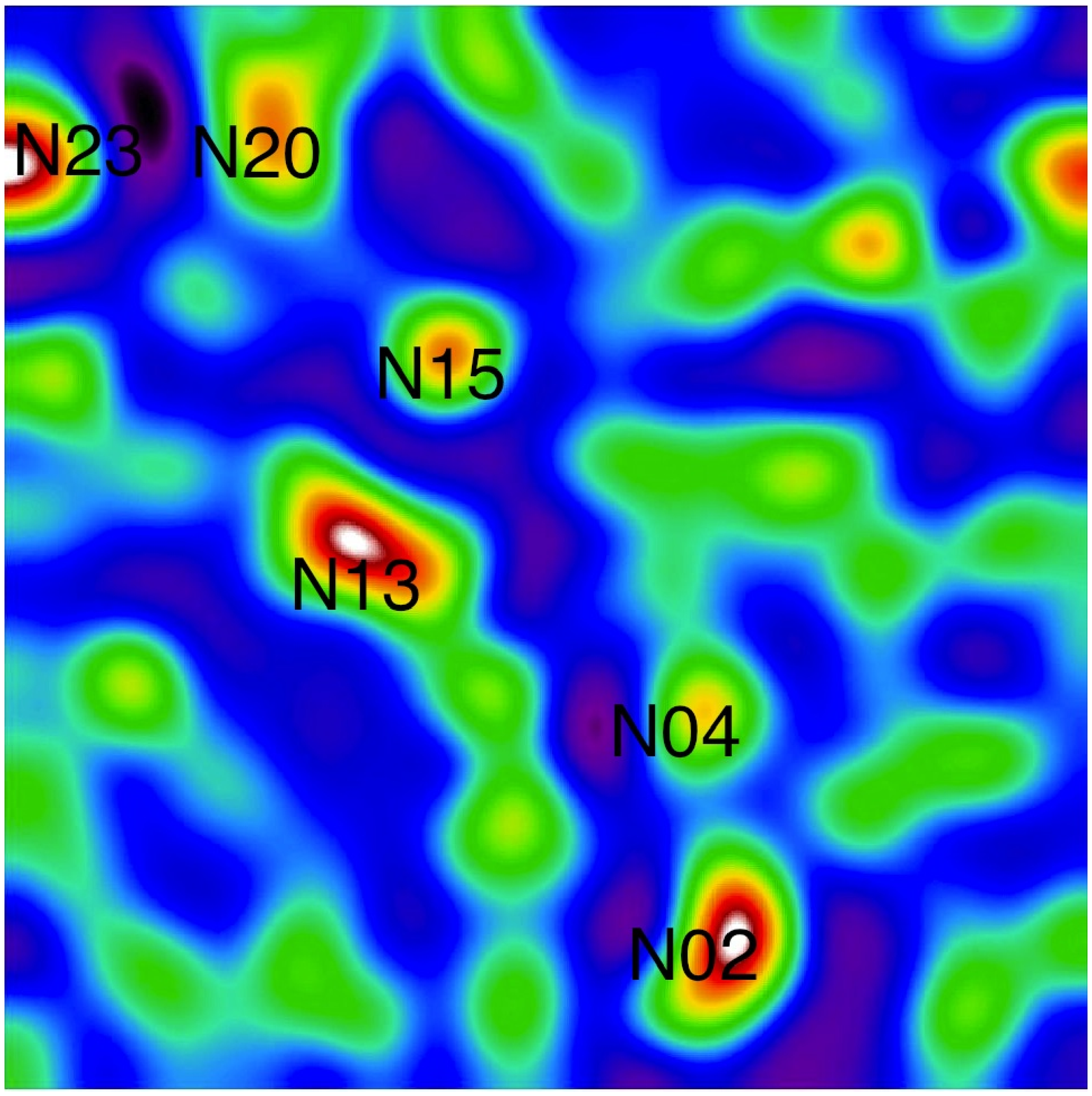}}\\
\vspace{1mm}
\resizebox{0.48\textwidth}{!}{\includegraphics*{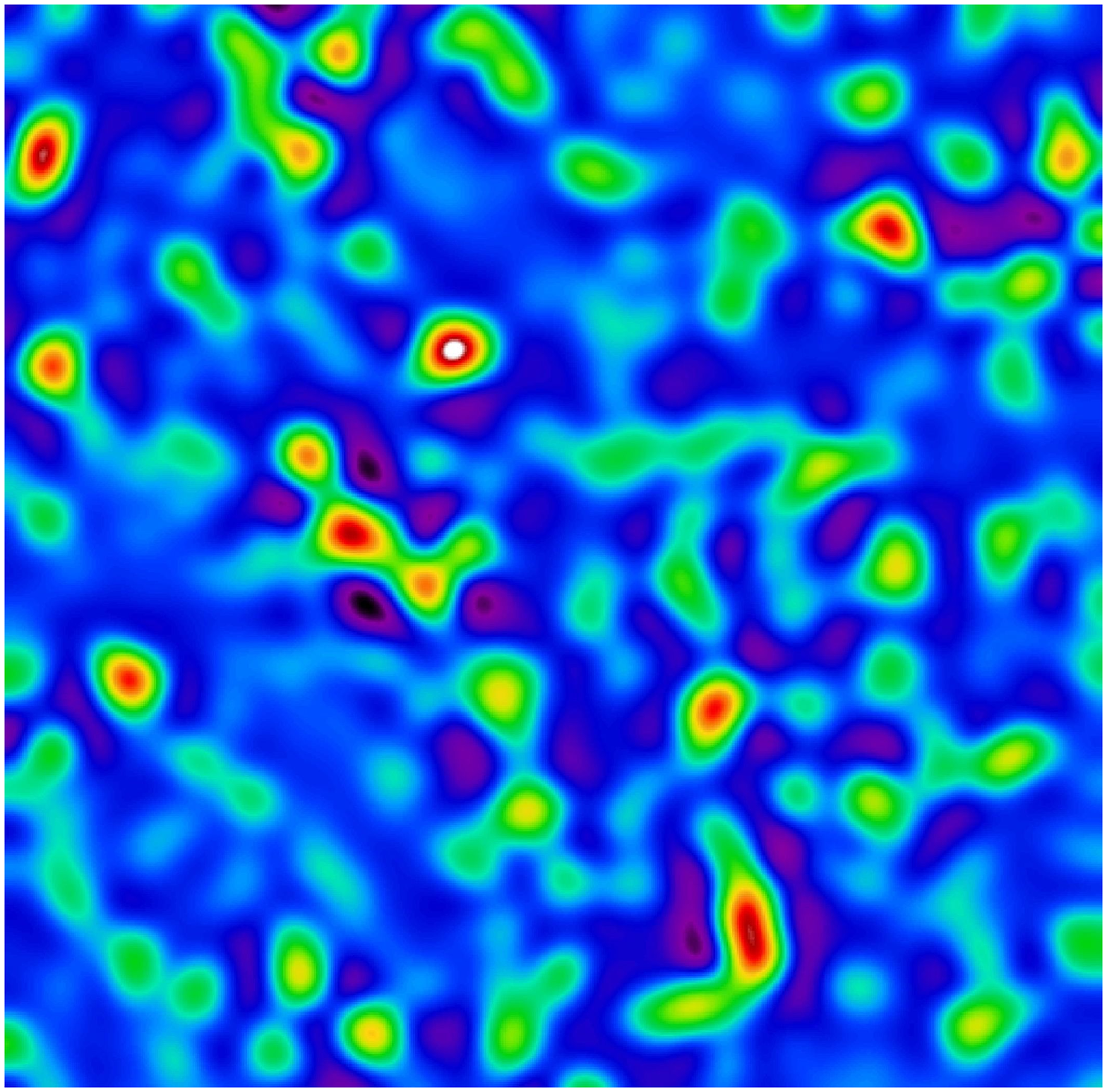}}
%\hspace{2mm} 
\resizebox{0.48\textwidth}{!}{\includegraphics*{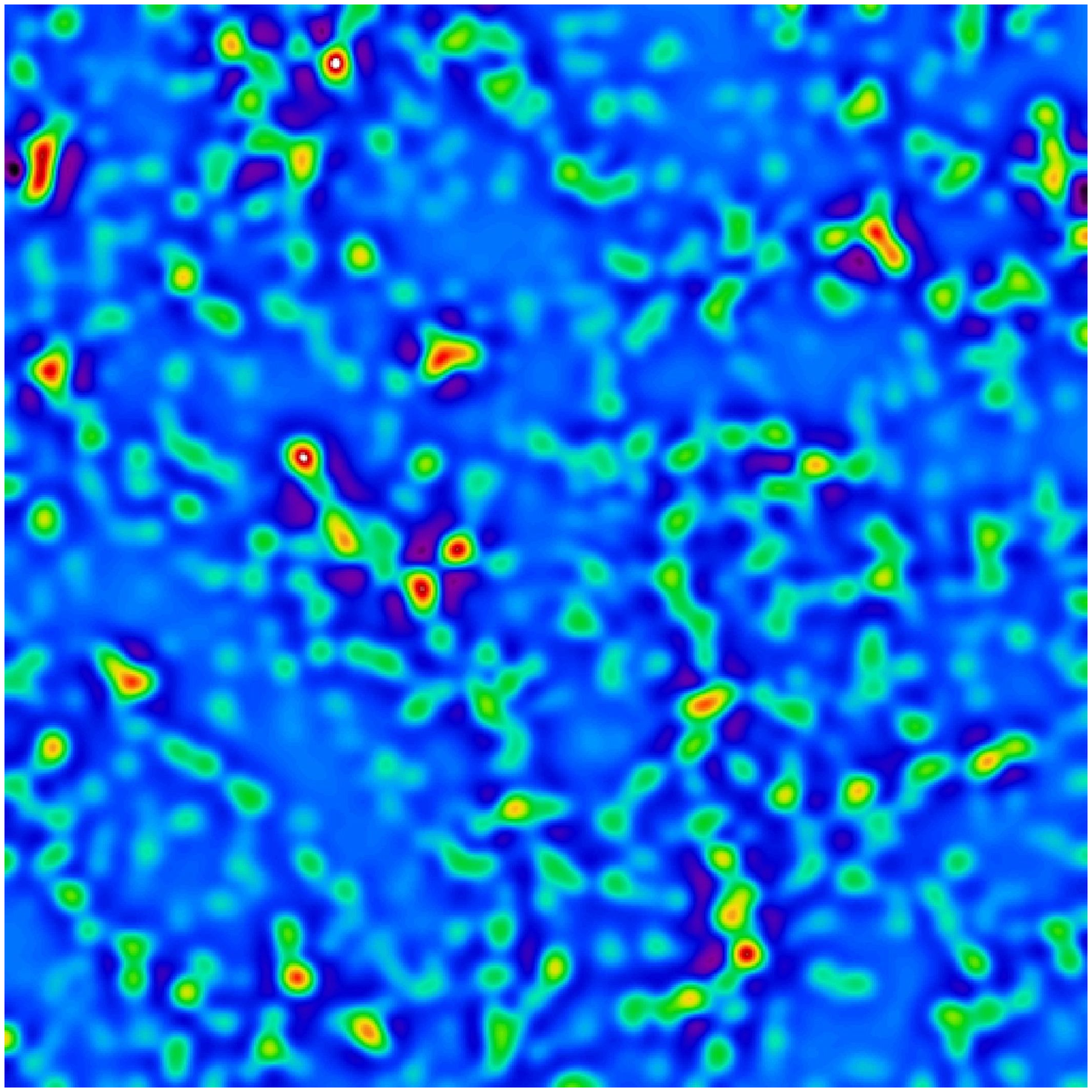}}
\end{minipage}
\caption{Upper left panel shows a two-dimensional rectangular region of  size $512\times 512$~Mpc (using Hubble constant $h = 0.8$), extracted from the   high-resolution luminosity density field of the  Northern equatorial wedge of   the SDSS. The density was   calculated using   Gaussian smoothing with the rms scale 0.8~Mpc.  The   observer is located at the lower left corner. Colour-coded density levels used in plotting are in the interval from 0 to 10 in mean density units with the highest value corresponding to the white colour, using the SAOImage program DS9 colour palette SLS. The upper right panel shows   the same  density field  as in the previous panel, but the phases   randomly   shifted. With shifting the phases of density waves some densities become negative, thus in this case colour codes are in the density interval $\pm 3.5$.  Middle and lower panels show the wavelets $w7$, $w6$, $w5$, and $w4$ of the  wedge.  The middle left panel shows  density waves of scales   approximately equal to half of the box size, each  following panel shows waves with half the length. Wavelet densities used in plotting have both negative and positive values, i.e., they correspond to under- and over-densities of matter, colour coding is linear. Supercluster numbers according to the catalogue by  \citet{Einasto:2003lh} are shown in the middle right panel $w6$.}  
\label{fig:sloan}
\end{figure}

For the present wavelet analysis we used the Northern equatorial wedge of 2.5 degrees thickness up to redshift $z=0.2$, see Fig.~\ref{fig:sloan}\footnote{This wavelet decomposition was first presented by \citet{Einasto:2006zr}.}. The high-resolution projected two-dimensional  luminosity density field was found using Gaussian smoothing with scale 0.8~Mpc.  Densities were found in a grid of bin size 1~Mpc. To reduce the wedge to unit thickness, density values were divided by the ratio of the thickness of the wedge at a particular distance from the observer, and at the mean distance.  We use a rectangular sub-region of the field suitable for the Fourier and wavelet analysis.

\subsection{ Fourier and wavelet decomposition of the SDSS density field}

The importance of the phase information in the cosmic web formation  has been understood long ago, as demonstrated by \citet{Coles:2000} by randomising phases of a simulated filamentary network.  To study the role of phase information in more detail, we extracted a rectangular region with the box size 512~Mpc, calculated for the Hubble constant $h = 0.8$, from the density field of the SDSS Northern equatorial slice. This region is shown in the upper left panel of Fig.~\ref{fig:sloan}.  We  Fourier-transformed the 2D density field and randomised phases of all Fourier components, and thereafter Fourier-transformed it back to see the resulting density field. A similar procedure has been applied also by \citet{Coles:2000}. The modified field has the same amplitudes of all wave-numbers $k$ as the original field, only the phases of waves are different. Results are shown in the right upper panel of Fig.~\ref{fig:sloan}.  We see that the whole structure of superclusters, filaments and voids has gone, the field is fully covered by tiny randomly spaced density enhancements. There are no clusters of galaxies in this picture, comparable in the luminosity to real clusters of galaxies.  Fourier phase randomising in numerical models by \citet{Chiang:2000} and \citet{Coles:2000} shows similar results.

This simple example shows the importance of the density perturbation phases  in the cosmic web  formation.

\subsection{Phase synchronisation and the supercluster structure}

Next we use the wavelet analysis to investigate the role of density waves of different scales. 
Figure~\ref{fig:sloan} shows wavelets 7 to 4 of the Northern rectangular region.  These wavelets characterise waves of length  about 256, 128, 64, and  32  Mpc, respectively (note that all scales in the Figure~\ref{fig:sloan} are given using Hubble constant $h=0.8$).  In wavelet figures both under- and  over-densities are shown. Extreme levels were chosen so that main features of the structure are well visible.

The middle left panel of Fig.~\ref{fig:sloan} shows the waves of length about 256 Mpc.  In its highest density regions there are three very rich superclusters: N20 from the list by \citet{Einasto:2003lh}, located in the upper  part of the figure, supercluster N13 (SCL126  from the list by \citet{Einasto:2001ff} in the Sloan Great Wall) near the centre, and supercluster N02 (SCL82) in the lower right part of the panel.

The next panel shows waves of scales about 128 Mpc. Here the most prominent features are superclusters N13 (SCL126) and N02 (SCL82), also the supercluster N23 (SCL155) in the upper left part of the panel is fairly strong, seen as a weak density peak already in the previous panel. In addition we see the supercluster N15 just above N13 near the minimum of the wave of 256 Mpc scale, and a number of poorer superclusters located mostly in voids defined by waves of larger size.

The lower left panel plots waves of scales about 64 Mpc. Here all superclusters seen on larger scales are also visible.  A large fraction of density enhancements are either situated just in the middle of low-density regions of the previous panel, or they divide massive superclusters into smaller subunits.  This property is repeated in the next panel.  Here the highest peaks are substructures of rich superclusters, and there are numerous smaller density enhancements (clusters) between the peaks of the previous panel.

When we compare the density waves of all scales,  we arrive at the conclusion that superclusters form in regions where  density waves of medium and large scales combine in {\em similar over-density phases}.  The larger  the scale of the wave where this coincidence takes place, the richer the supercluster.

Similarly, voids form in regions where  density waves of medium and large scales combine in {\em similar under-density phases}.  In large voids medium-scale perturbations generate a web of filamentary structures with knots; for a description of the formation of such a web see \citet{Bond:1996}. The influence of density perturbations of various scales on the  cosmic web formation  is discussed in more detail by \citet{Suhhonenko:2011}.

This simple analysis very clearly demonstrates the role of phase coupling (synchronisation) of density waves of different scales in the formation of the supercluster-void network.  A wavelet analysis of the full SDSS DR7 contiguous Northern region is in progress, results will be published in a separate paper.

\vspace{5mm}
\begin{figure}[ht]
\centering
\resizebox{0.45\textwidth}{!}{\includegraphics{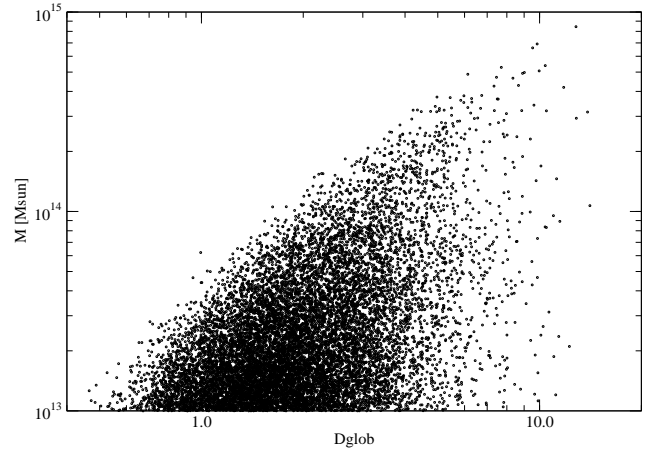}}
\caption{DF cluster masses of the model L256 as a function of the global density. Masses are given in solar units, densities in units of the mean density of the model.}
\label{fig:DFclust}
\end{figure}

\section{Discussion}

\subsection{The role of phase synchronisation in the formation of clusters and superclusters}
\begin{figure*}[ht]
\centering
\resizebox{0.9\textwidth}{!}{\includegraphics*{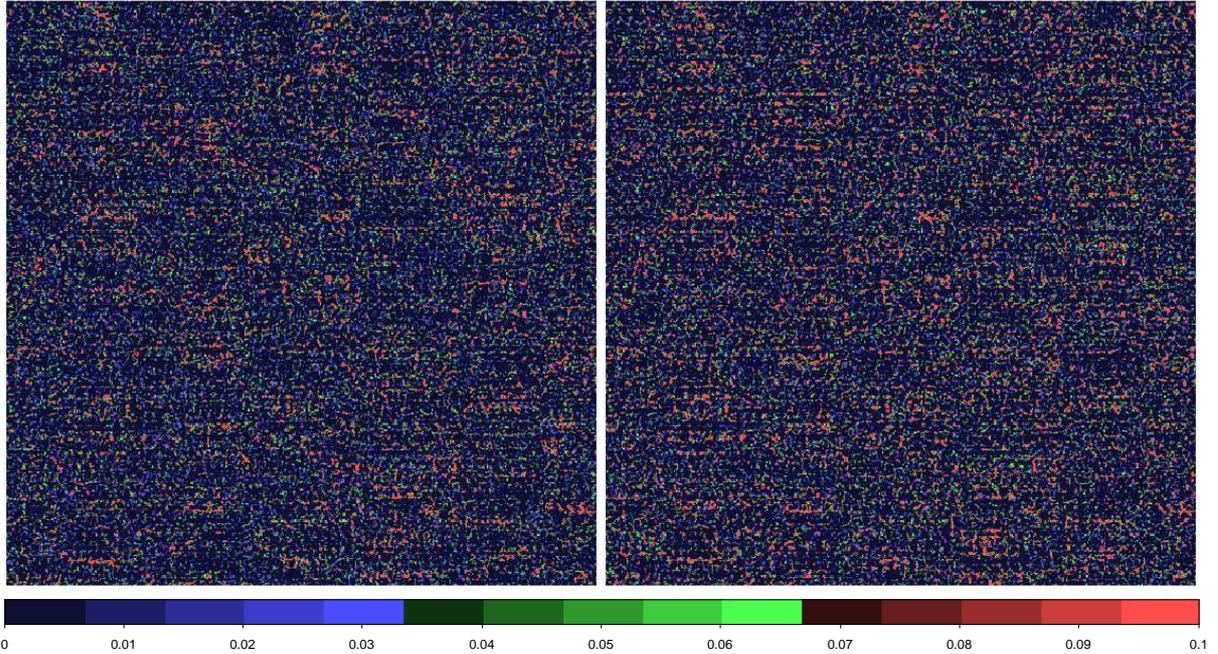}}
\caption{Wavelets $w1$ of models L256.256 and L256.008 at redshift $z=30$
are shown in the left and right panel, respectively, at coordinate $k=1$. Densities are expressed in linear scale, only over-density regions are shown. 
 }
\label{fig:L256w1}
\end{figure*}

The wavelet decomposition of SDSS data has shown that superclusters form in regions where  density waves of medium and large scales combine in  similar over-density phases. Now we shall consider the role of phase synchronisation in the formation of galaxy clusters.  

Our previous analysis has shown that clusters of galaxies (haloes in simulations) form in places where small density waves in a certain range of scales combine in similar over-density regions of waves. This is clearly seen as high level correlations of wavelets of levels 1, 2, and 3 in  Fig.~\ref{fig:correlate:fixed_z}, and close positions of maxima of wavelets $w4$ and $w5$, and positions of rich clusters in the high-resolution density field in Fig.~\ref{fig:sloan}.

To see the role of phase synchronisation in the formation of clusters we shall use the density field (DF) clusters of the simulation of box length $L=256$~\Mpc\ with resolution $512^3$ particles and cells, the model L256. We define DF clusters  as peaks of the high-resolution density field, calculated with the kernel of radius equal to the cell size of the simulation, the field $D_0$ (see Appendix A.3).  DF cluster positions were identified as grid coordinates $i,~j,~k$ of the local maxima of the field $D_0$.  The mass of the DF clusters, $M_{\mathrm{cl}}$, was found by counting local density values in cells within $\pm 4$ cells from the central one, i.e., within a box of half-length 2.5~\Mpc, centred on the peak. We express DF cluster masses in solar units using the masses of particles. This sample of DF clusters was used by \citet{Suhhonenko:2011} to find DF cluster mass distribution and cluster-defined void radii.  To investigate the growth of structures in the standard $\Lambda$CDM cosmogony, \citet{Ludlow:2010tg}  used a  slightly different algorithm to find density peaks in simulations.

The relationship between DF cluster masses and global densities is shown in Fig.~\ref{fig:DFclust}.  Global densities at the peak positions were found using the density field calculated with a kernel of radius 8~\Mpc, the field $D_4$.  The figure shows that there is an upper limit of masses of DF clusters for any given global density value.  Global densities calculated with a kernel of radius 8~\Mpc\ were used to define superclusters  \citep{Einasto:2003lh,Einasto:2006,Einasto:2007a,Liivamagi:2010}, using a density threshold $\simeq 4.5$ in mean density units.  We see that most massive DF clusters are located in superclusters.  As the global density decreases, the upper mass limit of DF clusters also decreases.  A similar tendency is found in SDSS clusters of galaxies by \citet{Einasto:2005a}: in high-density environment clusters have a mean luminosity a factor of $\simeq 5$ higher than in a low-density environment.  This is the result of phase synchronisation: the larger  the scale of density waves, where the maxima of waves of different scales have close positions, the larger are the masses of DF clusters. This also explains  why there are no very rich clusters in a low-density environment: amplification by large-scale density waves is needed to form a very massive cluster. This amplification occurs only in the regions we call superclusters, as seen also in the SDSS data in Fig.~\ref{fig:sloan}.

Figure~\ref{fig:DFclust} shows that at any given global density value there is a large range of masses of DF clusters.  This has a simple explanation.  Small-mass DF clusters are located far from the maxima of large-scale waves, seen in wavelets of higher order. 

We  conclude that in the formation of both small and large systems of galaxies the synchronisation of phases of density waves of different scales plays an important role.    Our experience shows that the wavelet analysis has a good diagnostic value in understanding the evolution of systems of galaxies of different richness.

\subsection{Phase synchronisation as a physical process}

Phase synchronisation between different scales is a physical process that 
requires a causal connection between different points in space. Because standard
inflation provides us with this connection for scales up to very large ones, 
much exceeding the present Hubble scale, this synchronisation can  already
 imprinted into the initial spectrum -- actually, this is the standard 
modern paradigm. However, it has not yet been proved if this initial 
synchronisation under the assumption of Gaussian statistics of 
perturbations in the linear regime is sufficient for the complete
explanation of the structure and evolution of the cosmic web. Thus, our 
investigation may also shed light on the actual statistics of the primordial
spectrum including its behaviour at large deviations from rms values. 
An alternative possibility might be some physical hydrodynamical processes 
acting before and during recombination. However, scales larger than the sound
horizon at recombination, $\approx 146$ Mpc according to the most recent
cosmological data by \citet{Jarosik:2010} remain unaffected by any such 
process.

Finally, larger scales may affect the evolution of smaller ones in the
recent epoch.
It is well known that density perturbations of large scales evolve almost linearly and do not change their positions. This behaviour was used by \citet{Kofman:1988} in the adhesion theory of the evolution of the large-scale structure of the Universe. Applying  Burgers equations, these authors  predicted the actual geometrical structure of the cosmic web skeleton, using the primordial (post-inflation) gravitation potential field.   Numerical simulations with the same initial conditions confirmed the correctness of the prediction. This  calculation shows that the cosmic web skeleton was created at a very early stage of the evolution of the Universe.   Our analysis has shown how this stability of the shape of the cosmic web skeleton  is expressed in wavelet terms.

To understand better the synchronisation of waves of different scales as a physical process, we used two  models of the L256 series. The first model has a full power spectrum, the model L256.256. The other model has a  power spectrum truncated at wave-number $k_{\mathrm{t}}$, so that the amplitude of the power spectrum on large scales is zero: $P(k) = 0$, if $k< k_{\mathrm{t}}$, wavelength $\lambda_{\mathrm{t}} = 2\pi/k_{\mathrm{t}} =8$~\Mpc, the model L256.008.  Data of these simulations are described by \citet{Suhhonenko:2011}.   

The power spectrum of density perturbations has the highest power at small scales. Accordingly, the influence of small-scale perturbations relative to large-scale perturbations is strongest in the early period of structure evolution.  For this reason the density fields and wavelets $w1$ at early epochs are almost identical for the full model L256.256, and for the model cut at small scales, $\lambda_{\mathrm{t}} = 8$~\Mpc, L256.008, see Fig.~\ref{fig:L256w1}.  Eventually, perturbations of larger scale start to affect the evolution. These perturbations amplify small-scale perturbations near maxima,  and suppress small-scale perturbations near minima.  In this way the growth of small-scale perturbations becomes non-linear.  Thereafter  still larger  perturbations amplify smaller perturbation near their maxima, and suppress smaller perturbations near their minima, and so on. The largest amplification (non-linearity) occurs in regions where maxima of perturbations of all scales happen to coincide.  In such a way the synchronisation of phases of waves of different scales occurs as a natural process.   The synchronisation of waves of different scales as a function of redshift $z$ is seen graphically in Fig.~7 of \citet{Einasto:2011}.

\section{Conclusions}

Main conclusions of the present paper can be formulated as follows.

\begin{enumerate}

\item The wavelet analysis has demonstrated a good diagnostic value for studying the evolution of galaxy systems  of various scales and masses.

\item In the formation of cosmic structures the synchronisation (coupling) of density waves of different scales plays an important role.

\item Positions of density maxima of waves of  large and medium scales practically do not change during the evolution. On smaller scales positions of density maxima change during the evolution, the changes are larger for waves with shorter wavelengths.

\item Superclusters are objects where density waves of  medium and large scales combine {\em in similar phases to  generate high-density regions}.

\item Voids are regions in space where  density waves of  medium and large scales combine {\em in similar  under-density phases}.

\item Clusters of galaxies are objects where density waves of small scales combine in similar over-density phases.

\item The larger is the scale of the highest phase synchronisation, the richer are the clusters and superclusters.

\end{enumerate}

\begin{acknowledgements}

We thank the referee for a constructive report that helped to improve the paper.
The present study was supported by the Estonian Science Foundation grants No.  7146,  and 8005, and by the Estonian Ministry for Education and Science grant SF0060067s08. It has also been supported by ICRAnet through a professorship for Jaan Einasto, and by the University of Valencia (Vicerrectorado de Investigaci\'on) through a visiting professorship for Enn Saar and by the Spanish MEC projects ``ALHAMBRA'' (AYA2006-14056) and ``PAU'' (CSD2007-00060), including FEDER contributions. J.E., I.S. and E.T.  thank Leibniz-Institut f\"ur Astrophysik Potsdam (using DFG-grant Mu 1020/15-1), where part of this study was performed.  J.E. thanks also the Aspen Center for Physics and the Johns Hopkins University for hospitality where this project was started and continued.  The simulation for the model L256 was calculated at the High Performance Computing Centre, University of Tartu. For plotting the density fields and wavelets we used the SAOImage DS9 program. A.A.S. acknowledges the RESCEU hospitality as a visiting professor. He was also partially supported by the Russian Foundation for Basic Research grant No. 11-02-00643 and by the Scientific Programme ``Astronomy'' of the Russian Academy of Sciences.

We thank the SDSS Team for the publicly available data releases.  Funding for the SDSS and SDSS-II has been provided by the Alfred P. Sloan Foundation, the Participating Institutions, the National Science Foundation, the U.S. Department of Energy, the National Aeronautics and Space Administration, the Japanese Monbukagakusho, the Max Planck Society, and the Higher Education Funding Council for England. The SDSS Web Site is \texttt{http://www.sdss.org/}.

The SDSS is managed by the Astrophysical Research Consortium for the Participating Institutions. The Participating Institutions are the American Museum of Natural History, Astrophysical Institute Potsdam, University of Basel, University of Cambridge, Case Western Reserve University, University of Chicago, Drexel University, Fermilab, the Institute for Advanced Study, the Japan Participation Group, Johns Hopkins University, the Joint Institute for Nuclear Astrophysics, the Kavli Institute for Particle Astrophysics and Cosmology, the Korean Scientist Group, the Chinese Academy of Sciences (LAMOST), Los Alamos National Laboratory, the Max-Planck-Institute for Astronomy (MPIA), the Max-Planck-Institute for Astrophysics (MPA), New Mexico State University, Ohio State University, University of Pittsburgh, University of Portsmouth, Princeton University, the United States Naval Observatory, and the University of Washington.

\end{acknowledgements}

\begin{appendix}

\section{Density field and wavelets}
\label{sec:DF}

\subsection{Density field}

To estimate the expected total luminosity of groups or single galaxies, we assume that
the luminosity functions derived for a representative volume can be applied also to
individual groups and galaxies.  Under this assumption the estimated total luminosity per one visible galaxy is
\begin{equation}
L_\mathrm{tot} = L_\mathrm{obs} W_d, 
\label{eq:ldens}
\end{equation}
where $L_\mathrm{obs}={L}_{\sun}10^{0.4\times (M_{\sun}-M)}$ is the
luminosity of the visible galaxy of absolute magnitude $M$ (in units
of the luminosity of the Sun, ${L}_{\sun}$), and 
\begin{equation}
W_d =  {\frac{\int_0^\infty L\,F(L)\mathrm{d}L}{\int_{L_1}^{L_2} L\,F(L)\mathrm{d}L}} 
\label{eq:weight2}
\end{equation}
is the luminous-density weight (the ratio of the expected total luminosity to
the expected luminosity in the visibility window). $L_1$ and $L_2$ are lower and
upper limit of the luminosity window, respectively.  In our calculations we
 adopted the absolute magnitude of the Sun in the $r$ filter $M_{\sun} =4.64$ \citep{Blanton:2007}.  

The $k$-correction for the SDSS galaxies is calculated using the KCORRECT algorithm (version v4.1.4) developed by  \citet{Blanton:2003a} and \citet{Blanton:2007}. Evolution correction $e$ has been calculated according to \citet{Blanton:2003}.  For details of the data reduction procedure see \citet{Tago:2010}.

In calculating of the total expected luminosity we used a double-power-law luminosity function with a smooth transition:
\begin{equation}
\phi (L) \mathrm{d}L \propto (L/L^{*})^\alpha (1 +
(L/L^{*})^\gamma)^{(\delta-\alpha)/\gamma}
\mathrm{d}(L/L^{*}), 
\label{eq:abell}
\end{equation}
where $\alpha$ is the exponent at low luminosities $(L/L^{*}) \ll 1$, $\delta$ is the exponent at high luminosities $(L/L^{*}) \gg 1$, $\gamma$ is a parameter that determines the speed of transition between the two power laws, and $L^{*}$ is the characteristic luminosity of the transition, similar to the characteristic luminosity of the Schechter function.  As demonstrated by  \citet{Tempel:2009}, the double-power-law luminosity function fits observed luminosity distribution of galaxies and groups of galaxies better than the Schechter function, in particular in the  high-luminosity end of the distribution.

\subsection{Kernel method}

To calculate a density field, 
we need to convert the spatial positions of galaxies $\mathbf{r}_i$ 
and their luminosities  $L_i$ into
spatial (luminosity) densities. The standard approach is to use kernel densities
%\citep{silverman86}:
\begin{equation}
    \rho(\mathbf{r}) = \sum_i K\left( \mathbf{r} - \mathbf{r}_i; a\right) L_i,
\end{equation}
where the sum is over all galaxies, and $K\left(\mathbf{r};
a\right)$ is a kernel function of a width $a$. Good kernels
for calculating densities on a spatial grid are generated by box splines
$B_J$. Box splines are local and they are interpolating on a grid:
\begin{equation}
    \sum_i B_J \left(x-i \right) = 1,
\end{equation}
for any $x$ and a small number of indices that give non-zero values for $B_J(x)$.

We use the popular $B_3$ spline function:
\begin{equation} %[
B_3(x)=\frac1{12}\left[|x-2|^3-4|x-1|^3+6|x|^3-4|x+1|^3+|x+2|^3\right].
\end{equation} %]
We define the (one-dimensional) $B_3$ box spline kernel $K_B^{(1)}$ of the width $a$ as
\begin{equation}
    K_B^{(1)}(x;a,\delta) = B_3(x/a)(\delta / a),
\end{equation}
where $\delta$ is the grid step. This kernel differs from zero only
in the interval $x\in[-2a,2a]$; it is close to a Gaussian with $\sigma=1$ in the
region $x\in[-a,a]$, so its effective width is $2a$ \citep[see, e.g.,][]{Saar:2009}.

The kernel preserves the
interpolation property exactly for all values of $a$ and $\delta$,
where the ratio $a/\delta$ is an integer. (This kernel can be used also if this ratio
is not an integer, 
and $a \gg \delta$; the kernel sums to 1 in this case, too, with a very small error).
This means that if we apply this kernel to $N$ points on a one-dimensional grid,
the sum of the densities over the grid is exactly $N$.

The three-dimensional kernel $K_B^{(3)}$
is given by the direct product of three one-dimensional kernels:
\begin{equation}
    K_B^{(3)}(\mathbf{r};a,\delta) \equiv K_3^{(1)}(x;a,\delta) K_3^{(1)}(y;a,\delta) K_3^{(1)}(z;a,\delta),
\end{equation}
where $\mathbf{r} \equiv \{x,y,z\}$. Although this is a direct product,
it is isotropic to a good degree \citep{Saar:2009}.

To calculate the high-resolution density
field, we use the  kernel of a scale equal to the cell size of the
simulation, $l_c = L_b/N_{\mathbf{grid}}$, where $L_b$ is the size of the simulation box, and
$N_{\mathbf{grid}}$ is the number of grid elements in one coordinate.

\subsection{Wavelets}

 We use the '\'a trous' wavelet transform \citep{Martinez:2002ye}. 
The algorithm of the wavelet transform works as follows. Let us have a data set $D$ (particles in simulations or luminosity weighted galaxies in SDSS data), located in a box of size $n\times n \times n$. The wavelet transform decomposes the data set  as a superposition of the form
\begin{equation}
D = D_J + \sum_{j=1}^J w_j,
\end{equation}
where $D_J$ is the smoothed version of the original data $D$, and $w_j$ represents the structure of $D$ at scale $2^j$, \citep[see][]{Starck:1998sy,Starck:2002lt}.  The wavelet decomposition output is $J$ three-dimensional density fields $D_j$ and wavelets $w_j$ of size  $n\times n \times n$. Following the traditional indexing convention, we denote density fields and wavelets of the finest scale as $j=1$. All density fields were calculated with the $B_3$ spline kernel.  The smoothed version of the original data, $D_J = D_0$, is the density field  found with  the  kernel of the scale, equal to the cell size of the simulation $l_c$.

We calculated wavelets of index $j$ by subtraction of density fields:
\begin{equation}
w_j = D_{j-1} - D_{j},
\label{wi}
\end{equation}
where  every higher level density field $D_j$ was calculated with kernel size twice larger than the previous level field $D_{j-1}$.  In such construction a wavelet of index $j$ corresponds to density waves between  scales $\Delta_{j-1}  = l_c \times 2^{j-1/2}$  and   $\Delta_{j}  =l_c\times 2^{j+1/2}$, i.e., diameters of kernels used in calculation of density fields $D_{j-1}$ and $D_j$.

\end{appendix}


\begin{thebibliography}{53}
\expandafter\ifx\csname natexlab\endcsname\relax\def\natexlab#1{#1}\fi

\bibitem[{{Abazajian} {et~al.}(2009){Abazajian}, {Adelman-McCarthy},
  {Ag{\"u}eros}, {Allam}, {Allende Prieto}, {An}, {Anderson}, {Anderson},
  {Annis}, {Bahcall}, {Bailer-Jones}, {Barentine}, {Bassett}, {Becker},
  {Beers}, {Bell}, {Belokurov}, {Berlind}, {Berman}, {Bernardi}, {Bickerton},
  {Bizyaev}, {Blakeslee}, {Blanton}, {Bochanski}, {Boroski}, {Brewington},
  {Brinchmann}, {Brinkmann}, {Brunner}, {Budav{\'a}ri}, {Carey}, {Carliles},
  {Carr}, {Castander}, {Cinabro}, {Connolly}, {Csabai}, {Cunha}, {Czarapata},
  {Davenport}, {de Haas}, {Dilday}, {Doi}, {Eisenstein}, {Evans}, {Evans},
  {Fan}, {Friedman}, {Frieman}, {Fukugita}, {G{\"a}nsicke}, {Gates},
  {Gillespie}, {Gilmore}, {Gonzalez}, {Gonzalez}, {Grebel}, {Gunn},
  {Gy{\"o}ry}, {Hall}, {Harding}, {Harris}, {Harvanek}, {Hawley}, {Hayes},
  {Heckman}, {Hendry}, {Hennessy}, {Hindsley}, {Hoblitt}, {Hogan}, {Hogg},
  {Holtzman}, {Hyde}, {Ichikawa}, {Ichikawa}, {Im}, {Ivezi{\'c}}, {Jester},
  {Jiang}, {Johnson}, {Jorgensen}, {Juri{\'c}}, {Kent}, {Kessler}, {Kleinman},
  {Knapp}, {Konishi}, {Kron}, {Krzesinski}, {Kuropatkin}, {Lampeitl},
  {Lebedeva}, {Lee}, {Lee}, {Leger}, {L{\'e}pine}, {Li}, {Lima}, {Lin}, {Long},
  {Loomis}, {Loveday}, {Lupton}, {Magnier}, {Malanushenko}, {Malanushenko},
  {Mandelbaum}, {Margon}, {Marriner}, {Mart{\'{\i}}nez-Delgado}, {Matsubara},
  {McGehee}, {McKay}, {Meiksin}, {Morrison}, {Mullally}, {Munn}, {Murphy},
  {Nash}, {Nebot}, {Neilsen}, {Newberg}, {Newman}, {Nichol}, {Nicinski},
  {Nieto-Santisteban}, {Nitta}, {Okamura}, {Oravetz}, {Ostriker}, {Owen},
  {Padmanabhan}, {Pan}, {Park}, {Pauls}, {Peoples}, {Percival}, {Pier}, {Pope},
  {Pourbaix}, {Price}, {Purger}, {Quinn}, {Raddick}, {Fiorentin}, {Richards},
  {Richmond}, {Riess}, {Rix}, {Rockosi}, {Sako}, {Schlegel}, {Schneider},
  {Scholz}, {Schreiber}, {Schwope}, {Seljak}, {Sesar}, {Sheldon}, {Shimasaku},
  {Sibley}, {Simmons}, {Sivarani}, {Smith}, {Smith}, {Smol{\v c}i{\'c}},
  {Snedden}, {Stebbins}, {Steinmetz}, {Stoughton}, {Strauss}, {Subba Rao},
  {Suto}, {Szalay}, {Szapudi}, {Szkody}, {Tanaka}, {Tegmark}, {Teodoro},
  {Thakar}, {Tremonti}, {Tucker}, {Uomoto}, {Vanden Berk}, {Vandenberg},
  {Vidrih}, {Vogeley}, {Voges}, {Vogt}, {Wadadekar}, {Watters}, {Weinberg},
  {West}, {White}, {Wilhite}, {Wonders}, {Yanny}, {Yocum}, {York}, {Zehavi},
  {Zibetti}, \& {Zucker}}]{Abazajian:2009kx}
{Abazajian}, K.~N., {Adelman-McCarthy}, J.~K., {Ag{\"u}eros}, M.~A., {et~al.}
  2009, \apjs, 182, 543

\bibitem[{{Albrecht} \& {Steinhardt}(1982)}]{Albrecht:1982}
{Albrecht}, A. \& {Steinhardt}, P.~J. 1982, Physical Review Letters, 48, 1220

\bibitem[{{Bertschinger}(1995)}]{Bertschinger:1995}
{Bertschinger}, E. 1995, ArXiv:astro-ph/9506070

\bibitem[{{Blanton} {et~al.}(2003{\natexlab{a}}){Blanton}, {Brinkmann},
  {Csabai}, {Doi}, {Eisenstein}, {Fukugita}, {Gunn}, {Hogg}, \&
  {Schlegel}}]{Blanton:2003}
{Blanton}, M.~R., {Brinkmann}, J., {Csabai}, I., {et~al.} 2003{\natexlab{a}},
  \aj, 125, 2348

\bibitem[{{Blanton} {et~al.}(2003{\natexlab{b}}){Blanton}, {Hogg}, {Bahcall},
  {Brinkmann}, {Britton}, {Connolly}, {Csabai}, {Fukugita}, {Loveday},
  {Meiksin}, {Munn}, {Nichol}, {Okamura}, {Quinn}, {Schneider}, {Shimasaku},
  {Strauss}, {Tegmark}, {Vogeley}, \& {Weinberg}}]{Blanton:2003a}
{Blanton}, M.~R., {Hogg}, D.~W., {Bahcall}, N.~A., {et~al.} 2003{\natexlab{b}},
  \apj, 592, 819

\bibitem[{{Blanton} \& {Roweis}(2007)}]{Blanton:2007}
{Blanton}, M.~R. \& {Roweis}, S. 2007, \aj, 133, 734

\bibitem[{{Bond} {et~al.}(1996){Bond}, {Kofman}, \& {Pogosyan}}]{Bond:1996}
{Bond}, J.~R., {Kofman}, L., \& {Pogosyan}, D. 1996, \nat, 380, 603

\bibitem[{{Chiang} \& {Coles}(2000)}]{Chiang:2000}
{Chiang}, L. \& {Coles}, P. 2000, \mnras, 311, 809

\bibitem[{{Chiang} {et~al.}(2002){Chiang}, {Coles}, \&
  {Naselsky}}]{Chiang:2002}
{Chiang}, L., {Coles}, P., \& {Naselsky}, P. 2002, \mnras, 337, 488

\bibitem[{{Coles}(2009)}]{Coles:2009}
{Coles}, P. 2009, in Lecture Notes in Physics, Berlin Springer Verlag, Vol.
  665, Data Analysis in Cosmology, ed. {V.~J.~Martinez, E.~Saar,
  E.~M.~Gonzales, \& M.~J.~Pons-Borderia }, 493

\bibitem[{{Coles} \& {Chiang}(2000)}]{Coles:2000}
{Coles}, P. \& {Chiang}, L. 2000, \nat, 406, 376

\bibitem[{{de Lapparent} {et~al.}(1986){de Lapparent}, {Geller}, \&
  {Huchra}}]{de-Lapparent:1986}
{de Lapparent}, V., {Geller}, M.~J., \& {Huchra}, J.~P. 1986, \apjl, 302, L1

\bibitem[{{Einasto}(2006{\natexlab{a}})}]{Einasto:2006mz}
{Einasto}, J. 2006{\natexlab{a}}, Commmunications of the Konkoly Observatory
  Hungary, 104, 163

\bibitem[{{Einasto}(2006{\natexlab{b}})}]{Einasto:2006zr}
{Einasto}, J. 2006{\natexlab{b}}, in Bernard's Cosmic Stories: From Primordial
  Fluctuations to the Birth of Stars and Galaxies

\bibitem[{{Einasto}(2009)}]{Einasto:2009ly}
{Einasto}, J. 2009, ArXiv:0906.5272

\bibitem[{{Einasto} {et~al.}(2006){Einasto}, {Einasto}, {Saar}, {Tago},
  {Liivam{\"a}gi}, {J{\~o}eveer}, {Suhhonenko}, {H{\"u}tsi}, {Jaaniste},
  {Hein{\"a}m{\"a}ki}, {M{\"u}ller}, {Knebe}, \& {Tucker}}]{Einasto:2006}
{Einasto}, J., {Einasto}, M., {Saar}, E., {et~al.} 2006, \aap, 459, L1

\bibitem[{{Einasto} {et~al.}(2007){Einasto}, {Einasto}, {Tago}, {Saar},
  {H{\"u}tsi}, {J{\~o}eveer}, {Liivam{\"a}gi}, {Suhhonenko}, {Jaaniste},
  {Hein{\"a}m{\"a}ki}, {M{\"u}ller}, {Knebe}, \& {Tucker}}]{Einasto:2007a}
{Einasto}, J., {Einasto}, M., {Tago}, E., {et~al.} 2007, \aap, 462, 811

\bibitem[{{Einasto} {et~al.}(2003){Einasto}, {H{\"u}tsi}, {Einasto}, {Saar},
  {Tucker}, {M{\"u}ller}, {Hein{\"a}m{\"a}ki}, \& {Allam}}]{Einasto:2003lh}
{Einasto}, J., {H{\"u}tsi}, G., {Einasto}, M., {et~al.} 2003, \aap, 405, 425

\bibitem[{{Einasto} {et~al.}(1980){Einasto}, {J{\~o}eveer}, \&
  {Saar}}]{Einasto:1980}
{Einasto}, J., {J{\~o}eveer}, M., \& {Saar}, E. 1980, \mnras, 193, 353

\bibitem[{{Einasto} {et~al.}(2011){Einasto}, {Suhhonenko}, {H{\"u}tsi}, {Saar},
  {Einasto}, {Liivam{\"a}gi}, {M{\"u}ller}, {Starobinsky}, {Tago}, \&
  {Tempel}}]{Einasto:2011}
{Einasto}, J., {Suhhonenko}, I., {H{\"u}tsi}, G., {et~al.} 2011,
  ArXiv:1105.2464

\bibitem[{{Einasto} {et~al.}(2005){Einasto}, {Tago}, {Einasto}, {Saar},
  {Suhhonenko}, {Hein{\"a}m{\"a}ki}, {H{\"u}tsi}, \& {Tucker}}]{Einasto:2005a}
{Einasto}, J., {Tago}, E., {Einasto}, M., {et~al.} 2005, \aap, 439, 45

\bibitem[{{Einasto} {et~al.}(2001){Einasto}, {Einasto}, {Tago}, {M{\"u}ller},
  \& {Andernach}}]{Einasto:2001ff}
{Einasto}, M., {Einasto}, J., {Tago}, E., {M{\"u}ller}, V., \& {Andernach}, H.
  2001, \aj, 122, 2222

\bibitem[{{Guth}(1981)}]{Guth:1981ys}
{Guth}, A.~H. 1981, \prd, 23, 347

\bibitem[{{Guth} \& {Pi}(1982)}]{Guth:1982}
{Guth}, A.~H. \& {Pi}, S. 1982, Physical Review Letters, 49, 1110

\bibitem[{{Hawking}(1982)}]{Hawking:1982}
{Hawking}, S.~W. 1982, Physics Letters B, 115, 295

\bibitem[{{Hikage} {et~al.}(2005){Hikage}, {Matsubara}, {Suto}, {Park},
  {Szalay}, \& {Brinkmann}}]{Hikage:2005bd}
{Hikage}, C., {Matsubara}, T., {Suto}, Y., {et~al.} 2005, \pasj, 57, 709

\bibitem[{{Jarosik} {et~al.}(2010){Jarosik}, {Bennett}, {Dunkley}, {Gold},
  {Greason}, {Halpern}, {Hill}, {Hinshaw}, {Kogut}, {Komatsu}, {Larson},
  {Limon}, {Meyer}, {Nolta}, {Odegard}, {Page}, {Smith}, {Spergel}, {Tucker},
  {Weiland}, {Wollack}, \& {Wright}}]{Jarosik:2010}
{Jarosik}, N., {Bennett}, C.~L., {Dunkley}, J., {et~al.} 2010, ArXiv:1001.4744

\bibitem[{{Jones}(2009)}]{Jones:2009dq}
{Jones}, B.~J.~T. 2009, in Lecture Notes in Physics, Berlin Springer Verlag,
  Vol. 665, Data Analysis in Cosmology, ed. {V.~J.~Martinez, E.~Saar,
  E.~M.~Gonzales, \& M.~J.~Pons-Borderia }, 3--50

\bibitem[{{Knebe} {et~al.}(2001){Knebe}, {Green}, \& {Binney}}]{Knebe:2001qa}
{Knebe}, A., {Green}, A., \& {Binney}, J. 2001, \mnras, 325, 845

\bibitem[{{Kofman} \& {Shandarin}(1988)}]{Kofman:1988}
{Kofman}, L.~A. \& {Shandarin}, S.~F. 1988, \nat, 334, 129

\bibitem[{{Liivam{\"a}gi} {et~al.}(2010){Liivam{\"a}gi}, {Tempel}, \&
  {Saar}}]{Liivamagi:2010}
{Liivam{\"a}gi}, L.~J., {Tempel}, E., \& {Saar}, E. 2010, ArXiv:1012.1989

\bibitem[{{Linde}(1982)}]{Linde:1982fr}
{Linde}, A.~D. 1982, Physics Letters B, 108, 389

\bibitem[{{Ludlow} \& {Porciani}(2010)}]{Ludlow:2010tg}
{Ludlow}, A.~D. \& {Porciani}, C. 2010, ArXiv:1011.2493

\bibitem[{{Mart{\'{\i}}nez} \& {Saar}(2002)}]{Martinez:2002ye}
{Mart{\'{\i}}nez}, V.~J. \& {Saar}, E. 2002, {Statistics of the Galaxy
  Distribution} (Chapman \& Hall/CRC)

\bibitem[{{Mart{\'{\i}}nez} {et~al.}(2005){Mart{\'{\i}}nez}, {Starck}, {Saar},
  {Donoho}, {Reynolds}, {de la Cruz}, \& {Paredes}}]{Martinez:2005kl}
{Mart{\'{\i}}nez}, V.~J., {Starck}, J., {Saar}, E., {et~al.} 2005, \apj, 634,
  744

\bibitem[{{Mukhanov} \& {Chibisov}(1981)}]{Mukhanov:1981}
{Mukhanov}, V.~F. \& {Chibisov}, G.~V. 1981, Soviet Journal of Experimental and
  Theoretical Physics Letters, 33, 532

\bibitem[{{Romeo} {et~al.}(2008){Romeo}, {Agertz}, {Moore}, \&
  {Stadel}}]{Romeo:2008}
{Romeo}, A.~B., {Agertz}, O., {Moore}, B., \& {Stadel}, J. 2008, \apj, 686, 1

\bibitem[{{Ryden} \& {Gramann}(1991)}]{Ryden:1991fk}
{Ryden}, B.~S. \& {Gramann}, M. 1991, \apjl, 383, L33

\bibitem[{{Saar}(2009)}]{Saar:2009}
{Saar}, E. 2009, in Lecture Notes in Physics, Berlin Springer Verlag, Vol. 665,
  Data Analysis in Cosmology, ed. {V.~J.~Martinez, E.~Saar, E.~M.~Gonzales, \&
  M.~J.~Pons-Borderia }, 523--563

\bibitem[{{Slezak} {et~al.}(1993){Slezak}, {de Lapparent}, \&
  {Bijaoui}}]{Slezak:1993nx}
{Slezak}, E., {de Lapparent}, V., \& {Bijaoui}, A. 1993, \apj, 409, 517

\bibitem[{{Springel}(2005)}]{Springel:2005a}
{Springel}, V. 2005, \mnras, 364, 1105

\bibitem[{{Starck} \& {Murtagh}(2002)}]{Starck:2002lt}
{Starck}, J. \& {Murtagh}, F. 2002, {Astronomical image and data analysis}, ed.
  {Starck, J.-L.~\& Murtagh, F.}

\bibitem[{{Starck} {et~al.}(1998){Starck}, {Murtagh}, \&
  {Bijaoui}}]{Starck:1998sy}
{Starck}, J., {Murtagh}, F.~D., \& {Bijaoui}, A. 1998, {Image Processing and
  Data Analysis}, ed. {Starck, J.-L., Murtagh, F.~D., \& Bijaoui, A.}

\bibitem[{{Starobinsky}(1979)}]{Starobinskii:1979vn}
{Starobinsky}, A.~A. 1979, Soviet Journal of Experimental and Theoretical
  Physics Letters, 30, 682

\bibitem[{{Starobinsky}(1980)}]{Starobinsky:1980rt}
{Starobinsky}, A.~A. 1980, Physics Letters B, 91, 99

\bibitem[{{Starobinsky}(1982)}]{Starobinsky:1982zr}
{Starobinsky}, A.~A. 1982, Physics Letters B, 117, 175

\bibitem[{{Suhhonenko} {et~al.}(2011){Suhhonenko}, {Einasto}, {Liivam\"agi},
  {Saar}, {Einasto}, {H\"utsi}, {M\"uller}, {Starobinsky}, {Tago}, \&
  {Tempel}}]{Suhhonenko:2011}
{Suhhonenko}, I., {Einasto}, J., {Liivam\"agi}, L.~J., {et~al.} 2011,
  ArXiv:1101.0123

\bibitem[{{Tago} {et~al.}(2010){Tago}, {Saar}, {Tempel}, {Einasto}, {Einasto},
  {Nurmi}, \& {Hein{\"a}m{\"a}ki}}]{Tago:2010}
{Tago}, E., {Saar}, E., {Tempel}, E., {et~al.} 2010, \aap, 514, A102

\bibitem[{{Tempel} {et~al.}(2009){Tempel}, {Einasto}, {Einasto}, {Saar}, \&
  {Tago}}]{Tempel:2009}
{Tempel}, E., {Einasto}, J., {Einasto}, M., {Saar}, E., \& {Tago}, E. 2009,
  \aap, 495, 37

\bibitem[{{Tempel} {et~al.}(2011){Tempel}, {Saar}, {Liivam{\"a}gi}, {Tamm},
  {Einasto}, {Einasto}, \& {M{\"u}ller}}]{Tempel:2011}
{Tempel}, E., {Saar}, E., {Liivam{\"a}gi}, L.~J., {et~al.} 2011, \aap, 529, A53

\bibitem[{{van de Weygaert} \& {Schaap}(2009)}]{van-de-Weygaert:2009bh}
{van de Weygaert}, R. \& {Schaap}, W. 2009, in Lecture Notes in Physics, Berlin
  Springer Verlag, Vol. 665, Data Analysis in Cosmology, ed.
  {V.~J.~Mart{\'{\i}}nez, E.~Saar, E.~Mart{\'{\i}}nez-Gonz{\'a}lez, \&
  M.-J.~Pons-Border{\'{\i}}a}, 291--413

\bibitem[{{Vielva} {et~al.}(2006){Vielva}, {Mart{\'{\i}}nez-Gonz{\'a}lez}, \&
  {Tucci}}]{Vielva:2006tg}
{Vielva}, P., {Mart{\'{\i}}nez-Gonz{\'a}lez}, E., \& {Tucci}, M. 2006, \mnras,
  365, 891

\bibitem[{{Zeldovich} {et~al.}(1982){Zeldovich}, {Einasto}, \&
  {Shandarin}}]{Zeldovich:1982}
{Zeldovich}, Y.~B., {Einasto}, J., \& {Shandarin}, S.~F. 1982, \nat, 300, 407

\end{thebibliography}
\end{document}